\documentstyle[12pt,aasms4]{article}

\lefthead{P. B. Ivanov $\&$ I. D. Novikov}
\righthead{A new model of tidally disrupted star}

\begin{document}

\title{A new model of a tidally disrupted star}

\author{Pavel B. Ivanov\altaffilmark{1}}
\affil{Theoretical Astrophysics Center, Juliane Maries Vej 30,
           DK-2100 Copenhagen \O, Denmark; pavel@tac.dk}
\and

\author{Igor D. Novikov\altaffilmark{1,2,3}}
\affil{
Theoretical Astrophysics Center, Juliane Maries Vej 30,
           DK-2100 Copenhagen \O, Denmark; novikov@tac.dk }

\altaffiltext{1}{Astro Space Center of 
P. N. Lebedev Physical Institute, 84/32 Profsoyuznaya Street,
           Moscow, 117810, Russia}
\altaffiltext{2}{University Observatory, Juliane Maries Vej 30,
           DK-2100 Copenhagen \O, Denmark}
\altaffiltext{3}{Nordita, Blegdamsvej 17, DK-2100 Copenhagen \O, Denmark}


\begin{abstract}
A new semi-analytical model of a star evolving in a tidal field
is proposed. The model is a generalization of the 
so-called 'affine' stellar
model. In our model
the star is composed of
elliptical shells with different parameters and different orientations, 
depending on
time and on the radial Lagrangian coordinate of the shell. The evolution
equations of this model are derived 
from the virial relations under
certain assumptions, and the integrals of motion are
identified.
It is shown that the evolution equations can
be deduced from a variational principle.
The evolution equations are solved numerically and compared quantitatively
with the results of 3D numerical computations of the tidal interaction of
a star with a supermassive black hole.
The comparison shows very good agreement between the  
main ``integral'' characteristics describing the tidal interaction event 
in our model and in the 3D computations. Our model is effectively 
a one-dimensional Lagrangian model from the point of view
of numerical computations,
and therefore it can be evolved numerically $10^{2}-10^{3}$
times faster than
the 3D approach allows. This makes our model well suited
for intensive calculations  covering the whole parameter space
of the problem.

\end{abstract}




\section{Introduction}

Starting from the seminal paper by Roche the problem of the
tidal influence of a
gravitating source on a satellite has been addressed by numerous researchers.
More recently, interest in this problem has been raised by a paper of
Hills (Hills 1975), who proposed tidal disruption processes as the main 
processes of fueling of QSO's and AGN's. From the point of view
of the astrophysics of
QSO's and AGN's there are several approaches to that problem. Firstly
one can consider the tidal interaction event as an elementary process in
the complicated astrophysical environment of a supermassive black hole,
presumably situated in the cores of QSO's and AGN's. Then one could find
the main evolutionary characteristics of such a system and its average
luminosity,
taking into account additional gas dynamical and stellar dynamical processes
occurring in the cores (e.g. Hills, 1975, Frank $\&$ Rees 1976, Young et al
1977; Young 1977; Hills 1978; Frank 1979; Gurzadian $\&$ Ozernoi 1980;
Lacy et al 1982;  Illarionov $\&$ Romanova 1986a;
Illarionov $\&$ Romanova 1986b; Dokuchaev 1991; Beloborodov et al 1992;
Roos 1992; Syer $\&$ Ulmer 1999; Magorrian $\&$ Tremaine 1999).
One can also consider the evolution of remnants of a single tidal stripping or
tidal disruption
event and find the characteristic luminosity change of the 
object due to the accretion of the remnants through an accretion disk or
quasi-spherical configuration onto the  central black hole (e.g. Lacy et all 1982;
Rees 1988; Evans $\&$ Kochanek 1989; Cannizzo et all 1990;
Roos 1992; Kochanek 1994; Ulmer et al 1998; Kim et al 1999; 
Syer $\&$ Ulmer 1999; Ulmer 1999).

On the other hand it is very important to understand quantitatively the
main characteristics of the tidal encounter itself, and a lot of
of work has been devoted to the physical processes
occurring in a star during its fly-by around a black hole. The papers
on that subject could be classified by the different 
stellar models used in the calculations.
The simplest possible approach to the problem uses 
an incompressible model of the star. Thus one can reduce 
the complicated hydrodynamical nonlinear partial differential equation
governing the evolution of the stellar gas to a set of ordinary differential
equations, which are easy to analyze by analytical and 
numerical means. The study of incompressible models
has been performed for Newtonian and relativistic tidal fields and
different kinds of orbits of the star 
(e.g. Nduka, 1971; Fishbone, 1973;
Mashhoon, 1975; Luminet $\&$ Carter 1986; 
Kosovichev $\&$ Novikov, 1992). However, this approach
is highly unrealistic, since effects
determined by the compressibility of the star can play a major role during
the tidal disruption event (e.g. Carter $\&$ Luminet 1982).

A significant step forward was made by Lattimer and Schramm (Lattimer $\&$
Schramm 1976) and by Carter and Luminet (Carter $\&$ Luminet 1983, 1985)
who proposed the so-called affine model of the tidally disrupted star, which
allows for the
compressibility of the stellar gas. In this model the law of
time evolution of different elements of the star is defined in terms of
some spatially uniform $3\times 3$ matrix ${\bf Q}(t)$:
$$x^{i}=Q^{i}_{j}(t)x^{j}_{0}, $$   
where $x^{i}$ are the components of the position vector of a gas element,
$x^{j}_{0}$ are the components of the position vector in some reference
state (say, before the tidal field ``is switched on''), 
and summation over repeated indices is assumed. Then one can find the evolution
equations for the matrix elements from the so-called virial relations written for
the whole star. The affine model has successfully been applied to the problem of
tidal interaction and tidal disruption of a star by a supermassive black hole
during close encounters (e.g. Carter $\&$ Luminet 1983, 1985; Luminet $\&$ Mark 1985;
Luminet $\&$ Carter 1986; Luminet $\&$ Pichon 1986; Novikov et al 1992;
Diener et al 1995). Lai, Rasio and Shapiro used the same model for an approximate 
treatment of an isolated rotating star, as well as for a star in a binary system
(e.g. Lai et al 1994; Lai $\&$ Shapiro 1995).

Recent progress in numerical simulations has allowed researchers
to perform direct 3D simulations
of the tidal interaction and tidal disruption events. 
The first SPH simulations were run in the beginning of
eighties by Nolthenius and Katz, and by Bicknell and Gingold, although the number of particles
in these simulations was too small to be representative (Nolthenius $\&$ Katz 1982, 1983;
Bicknell $\&$ Gingold 1983). In the following decades,
SPH simulations were improved both by
increasing of the number of particles, and by using of more complicated stellar
models (e.g. Evans $\&$ Kochanek 1989; Laguna et al 1993; Laguna 1994; Fulbright et al 1995;
Ayal et al 2000). Three dimensional finite difference simulations were done
by Khokhlov,
Novikov and Pethick for a polytropic star in a Newtonian tidal field (Khokhlov et al 1993a,b,
hereafter Kh a,b), by Frolov, Khokhlov, Novikov and Pethick
for a white dwarf (Frolov et al 1994), and by Diener, Frolov, Khokhlov, Novikov and Pethick
for a polytropic star in the tidal field of a Kerr black hole (Diener et al 1997).  An interesting
attempt to combine the affine model and a simple version of 3D finite difference
hydrodynamics has been made by
Mark, Lioure and Bonazzola (Mark et al 1996).

Although the 3D simulations promise the most direct and thoughtful approach to the problem, they        
are still very time consuming. 
All in all, less than one hundred 
different sets of
values of the problem parameters have been tested with numerical experiments, 
and due to very poor statistics
these experiments cannot be used to characterize the general properties of the tidal encounters
for a broad range of available parameters. There is another, more fundamental difficulty connected
with the 3D simulations. The complexity of 3D hydrodynamical flows makes the interpretation of 
the results of numerical work increasingly difficult. The situation is reminiscent of a 
real physical experiment, and a simple 'reference' model of the tidally disrupted     
star would be very welcome
in order to interpret the results of the numerical simulations. 
On the other hand, the astrophysics 
of AGN's and QSO's requires a rather rough description of a single tidal encounter, and only a few
'averaged'  quantities such as e.g. the amount of mass lost by the star during the tidal interaction,
or the amount of energy deposited in the star by the tidal forces are of interest from the
astrophysical viewpoint.  

In this paper we  propose a new, semi-analytical model of the tidally interacting or 
tidally disrupted star which could 
be used for intensive calculations covering the whole parameter space of the problem, and also
as a 'reference' model for 3D simulations. Our model is a straightforward generalization of the
affine model. However, in contrast to the affine model, the different layers of the star evolve
differently in our model, and are connected to each other by a force determined by pressure. 
This allows us to employ our
model for calculation of quantities 
such as the loss of mass from the star after a fly-by over a black hole
without complete disruption, which cannot be calculated
in the affine approximation. Instead of the position matrix ${\bf Q}(t)$ of the affine model, we
use the position matrix ${\bf T}(t, r_{0})$, which depends not only on time, but in addition
on the value
$r_{0}$ of the 'reference' vector $x_{0}^{i}$ (obviously the radius $r_{0}$ plays the role of a 
Lagrangian coordinate, 
so we will later call it the Lagrangian radius). Thus, in our model the star consists
of elliptical shells which are
composed of all elements of the star with a given Lagrangian radius $r_{0}$. 
The evolution of the shell depends on the Lagrangian radius, and therefore the shells have
different ratios between their 
major axes and different rotation angles with respect to a (locally inertial) coordinate
frame centered on the star's center of mass, for the different values of the Lagrangian radius.
The evolution equations of our model follow from the virial relations written for each shell
(see e.g. Chandrasekhar 1969, hereafter Ch). Unlike
the affine model the virial relations written for a shell inside the star must
contain surface terms, and these surface terms lead to interactions between
shells with different Lagrangian radii, and therefore to the propagation of a
disturbance through the star. In fact, the evolution equations are of hyperbolic type, and
the disturbance induced by a tidal field propagates over the star as a non-linear sound
wave. We derive the evolution equations of our model in the next Section using certain approximations
for the pressure terms, and the terms describing the self-gravity of the star. In the simplest formulation
of our model the interaction between the shells depends only on their relative volumes, and
therefore the shells are allowed to intersect each other. Therefore the position matrix ${\bf T}$ has
no direct physical meaning in such a case, and one should only use 
quantities averaged over many shells 
in order to infer physical information (such as e.g. the energy and the angular momentum
contained inside some part of the star, the components of the quadrupole moment tensor for that
part, the amount of mass lost by the star, 
and so on). We show that the  energy and the angular momentum are well
defined in our model, and derive the law of evolution of these quantities due to the presence of a tidal
field. We also show that the circulation of velocity of a gas element over the shell is exactly 
conserved even in the presence of the tidal field. Then we apply our model to the simplest problem, 
the parabolic fly-by of a polytropic star around a source of Newtonian gravity, and numerically calculate
the evolution of the quantities characterizing the star during the fly-by. We compare our results with
 3D finite difference simulations 
of Kh a,b for the same problem and the same
parameters, 
and find very good agreement. 
Additionally we compare our model with results of
SPH simulations, calculations based on the affine model and results from the linear theory
of tidal perturbations (Press $\&$ Teukolsky 1977; Lee $\&$ Ostriker 1986).  
Then we calculate the energy deposited in the star, its angular
momentum and the amount of mass lost by the star
as a function of the pericentric separation between the star 
and the center of gravity. As it will be clear for the results in Section 3,
our model gives a better agreement with the results of the 3D simulations than the affine
model.

We use a rather unusual summation convention assuming 
that summation is performed over all indices appearing
in our expressions more than once, but summation is not performed if indices are enclosed
in brackets. Bold letters represent matrices in abstract form. All indices 
can be raised or lowered with help of the Kronecker
delta symbol, but nevertheless we distinguish between
the upper and lower indices in order to enumerate 
the rows and columns of matrices, respectively. Therefore, 
the expression $A^{k}_{i}B^{l}_{i}=C^{kl}$ 
means ${\bf AB}^{T}={\bf C}$, and $A^{i}_{k}B^{i}_{l}=C_{kl}$ means ${\bf A}^{T}{\bf B}={\bf C}$
(here $T$ stands for the transpose of a matrix).

Finally, we would like to list the main approximations made in the derivation of
the dynamical equations of our model. 

1) We assume that the star is composed of elliptical shells, and the shells are not deformed
during the evolution of the star in a tidal field.

2) We calculate the self-gravity of the star in a simplified manner. Namely, in order to calculate 
the force of gravity acting on some particular shell, we neglect the contribution of the star's mass
concentrated in the outer (with respect to that shell) layers of the star. It is also assumed that the
gravitational force determined by the inner layers is equivalent to the gravitational force of a
uniform density ellipsoid inserted in this shell.

3) We assume a polytropic equation of state of the stellar gas.

4) We use an ``averaged'' density and an ``averaged'' pressure instead of the exact quantities. 
These averaged quantities depend only on time and the Lagrangian radius of the star.

The approximations 1,2 are essential for our model. The approximations 3,4 can be relaxed in a more
advanced variant of the model.

\section{Building up of the model}

As was mentioned in the Introduction, we divide the star into 
a set of elliptical shells. Each shell consists of the gas 
elements which had the same distance from the center of the star
in the unperturbed spherical state.
The initial Cartesian coordinates of the gas elements of
the star  $x_{0}^{i}$ play the role
of Lagrangian coordinates in the course of the star's evolution under
the influence of a tidal field,
and hereafter we simply call them the Lagrangian
coordinates. We assume that the star layers corresponding to
the same Lagrangian radius $r_{0}=\sqrt{x_{0i}x^{i}_{0}}$ 
always keep
the elliptical form.  
The parameters of the shells
are different for the different values of $r_{0}$, and  
evolve with
time according to some dynamical equations, which are derived
below. Let us consider
the Eulerian coordinates of the gas elements $x^{i}$ with respect to 
some inertial reference system centered at the star's geometrical center.
From our discussion it follows that the law of transformation between
the Lagrangian and the Eulerian coordinates can be written as
$$x^{i}=T^{i}_{j}(t, r_{0})e_{0}^{j}, \eqno 1$$
where $e_{0}^{i}=x^{i}_{0}/r_{0}$, and $e_{0i}e^{i}_{0}=1$.
We also introduce the matrix
${\bf S}$ which is the inverse of the matrix ${\bf T}$.

The position matrix ${\bf T}$ and its inverse
${\bf S}$ can be represented as a product of
two rotational matrices ${\bf A}$ and ${\bf E}$, and a diagonal 
matrix ${\bf B}$:
$$T^{i}_{j}=A^{i}_{l}B^{l}_{m}E^{m}_{j}=a_{l}A^{i}_{l}E^{l}_{j}, 
\quad S^{i}_{j}=a_{l}^{-1}A^{j}_{l}E^{l}_{i}, \eqno 2$$
where $B^{l}_{m}=a_{(l)}\delta^{(l)}_{m}$, and $a_{l}$ are the principal
axes of the elliptical shell. Clearly, the matrices   ${\bf A}$ and 
${\bf E}$ describe the rotation of the principal axes of the shell
with respect to the reference frames in the Eulerian and 
Lagrangian spaces, respectively.
For our purpose it is useful to define several quantities
connected with the matrices   ${\bf T}$ and
${\bf S}$, namely the determinant $g$ of the position matrix:
$$g=|{\bf T}|=a_{1}a_{2}a_{3}, \eqno 3$$
the Jacobian $D=|{\partial x^{i} \over \partial x_{0}^{j}}|$ of the
mapping between the Lagrangian and Eulerian spaces:
$$D(x_{0}^{i})={ge_{0l}e_{n0}R^{ln} \over
r_{0}^{2}}, \eqno 4$$
where the symmetric matrix $R^{ln}$ determines the local shear and change of 
volume of the neighboring shells:
$$R^{ln}={1\over 2}(S^{l}_{m}(T^{m}_{n})\prime + S^{n}_{m}(T^{m}_{l})\prime), $$ 
and the prime stands for the differentiation with respect to $r_{0}$.
We also use the ``averaged'' Jacobian 
$$\bar D={1\over 4\pi}\int d\Omega D={dg\over dr_{0}^{3}}, \eqno 5$$
where the integration is performed over a unit sphere in the Lagrangian
space and $d\Omega$ is the elementary solid angle. 
Eq. (4)  immediately gives the law of evolution of the
gas density $\rho$ in our model. Taking into account
the law of mass conservation we have
$$\rho(t, x^{i})=\rho_{0}(r_{0})/D, \eqno 6$$
where $\rho_{0}(r_{0})$ is the density distribution 
in the unperturbed star. The ``averaged''
density $\bar \rho$ is defined analogously to (6), but with the
``averaged'' Jacobian (5)
$$\bar \rho (t, r_{0})=\rho_{0}/ {\bar D}=
{3\over 4\pi}{dM\over dg}, \eqno 7$$
where the mass differential $dM=4\pi\rho_{0}r_{0}^{2}dr_{0}$.
    
Of course, the transformation
law (1) is incompatible with the exact hydrodynamical equations
of motion of a perturbed star, and we must introduce some
reasonable approximations which allow us to reduce the
equations of motion to a dynamical equation for the matrix
components $T_{i}^{j}$. As a starting point of our analysis we use
the integral consequences of the exact equations of motion, namely
the equation of energy conservation and the so-called virial relations    
(see e. g. Ch, p. 20). In the adiabatic approximation
the energy equation
has the form
$${d\over dt}\lbrace \int d^3x(\rho v^{2}/2 +\epsilon)+{\mathcal P}\rbrace
=-\int dS_{i}(pv^{i})+\int d^3x(\rho C_{ij}v^{i}x^{j}), \eqno 8$$
and the virial relations are
$${d\over dt}\int d^3x(\rho x^{k}v^{i})=\int d^3x (\rho v^{k}v^{i})
+\delta^{ki}\int d^3 x p $$
$$-\int dS_{i}(x^{k}p)+{\mathcal P}^{ki}+
\int d^3x (\rho C^{i}_{j}x^{k}x^{j}). \eqno 9$$
Here $v^{i}$ is the
velocity of the gas element, $p$ is the pressure and $\epsilon$ is
the energy density per unit volume. 
The matrix $C^{i}_{j}$ represents the tidal tensor, and 
therefore it is symmetric and traceless. The potential energy
${\mathcal P}$ and the potential-energy tensor ${\mathcal P}^{ki}$
are
$${\mathcal P}=-{1\over 2}\int d^3x \int d^3x_{1} \rho(x^{i})
\rho(x_{1}^{i}){1\over |\vec x - \vec x_{1}|}, \eqno 10$$
and
$${\mathcal P}^{ki}=-{1\over 2}\int d^3x \int d^3x_{1} \rho(x^{i})
\rho(x_{1}^{i})
{(x^{k}-x^{k}_{1})(x^{i}-x_{1}^{i})
\over |\vec x - \vec x_{1}|}. \eqno 11$$
Obviously, ${\mathcal P}^{ki}$ is a symmetric matrix, and 
the relation
$${\mathcal P}={\mathcal P}^{kk} \eqno 12$$
holds. The volume integration in eqs. (8,9) and (10,11)
is performed over the
volume surrounded by a surface $r_{0}=$const, and the surface integration
is performed over that surface
\footnote{Strictly speaking we must extend the integration in the
inner integral in eqs. (10,11) over the whole volume of the
star. However in our approximation
the outer part of the star does not influence 
gravitationally the inner part of the star, 
and that part of the integrals 
can be omitted.}.

One can try to calculate the integrals containing the pressure and the energy
density (the thermal terms) in eqs (8,9), and the potential energy
tensor and the potential energy (the gravitational terms) directly,
using the transformation law (1), the density distribution (6), and
the adiabatic condition. However this approach leads to rather complicated
expressions for the thermal terms.
We want to construct our model in the simplest possible way, and
therefore we make several additional 
approximations with respect to these terms. 
For the thermal terms we use an ``averaged'' pressure 
$\bar p(t,r_{0})$, and an ``averaged'' energy density 
$\bar \epsilon (t, r_{0})$ instead of the exact quantities  
$p(t,x^{i})$ and $\epsilon (t, x^{i})$. That approximation allows us
to represent the thermal terms in eqs (8,9) in a very simple form
$$\delta^{ki}\int d^3 x p-\int dS_{i}(x^{k}p)={4\pi \over 3}\delta^{ki}
\int_{0}^{\bar p(r_{0})}g d\bar p, \eqno 13$$
$$\int dS_{i}(pv^{i})=4\pi H(t, r_{0})\bar p(t, r_{0}), \eqno 14$$
where the expansion rate is
$$H={1\over 3}S^{l}_{i}\dot T^{i}_{l}={1\over 3}({\dot a_{1} \over a_{1}}  
+{\dot a_{2} \over a_{2}}+{\dot a_{3} \over a_{3}}), \eqno 15$$
and the dot stands for time differentiation.
In our approximation the pressure force acting on the shell from the side of
the neighboring shells depends on the relative values of the shell volumes, and
does not depend on the orientation of the shells with respect to each other.
Therefore, the shells can intersect each other, and the interpretation of the 
position matrix $\bf {T}$ as describing the Eulerian positions of the gas elements
is rather ambiguous. As we have already mentioned in the Introduction
quantities are only meaningful
in this approximation
when averaged over many shells. In order to obtain the physical interpretation of
the position matrix $\bf {T}$ one should use a more complicated model,
where the pressure force depends on the relative orientation of the shells 
(see also Discussion).

In order to calculate an ``averaged'' potential energy tensor 
$\bar {\mathcal P}^{ik}$ we assume that the gravitational force acting
on the gas near the shell with some Lagrangian radius $r_{0}$ is equivalent
to the gravitational force of a uniform density ellipsoid with a mass equal to the part of
the star's mass within that shell. The principal
axes of that ellipsoid coincide with the principal axes of the shell,
and the density is averaged over the volume enclosed in the shell.
Under this assumption the ``averaged'' potential energy tensor 
$\bar {\mathcal P}^{ik}$ has the form
$$\bar {\mathcal P}^{ik}= -{1\over 2}\int dM GMA^{i}_{j}A^{k}_{j}
{a_{j}^{2}D_{j}\over g}, \eqno 16$$
and the averaged potential energy 
$\bar {\mathcal P}=  \bar {\mathcal P}^{kk}$ is
$$\bar {\mathcal P}= -{1\over 2}\int dM GM
{a_{j}^{2}D_{j}\over g}. \eqno 17$$
Here we use the mass $M=4\pi \int^{r_{0}}_{0}\rho_{0}(r_{1})dr_{1}$ 
of the gas inside the shell of the radius $r_{0}$ 
as a new Lagrangian coordinate instead of $r_{0}$. The dimensionless
quantities $D_{j}$ have been described by e. g. Ch, p. 41.
They have the form:
$$D_{j}=g\int^{\infty}_{0}{du\over \Delta (a_{j}^{2}+u)}, \eqno 18$$
where 
$$\Delta=\sqrt{(a_{1}^{2}+u)(a_{2}^{2}+u)(a_{3}^{2}+u)}.$$
These quantities obey very useful relations
$${a_{j}^{2}D_{j}\over g}=\int^{\infty}_{0}{du\over \Delta}, \eqno 19$$
and
$$\sum_{i=1}^{3} D_{i}=2.$$

Now we can substitute eqs (13,14) and (16,17) into eqs (8,9), and
perform the integration in the other terms with the help of the
law of mass conservation: $d^{3}x\rho=d^{3}x_{0}\rho_{0}(r_{0})$.
From the energy equation we obtain
$${d\over dt}\lbrace \int dM {\dot T^{i}_{n}\dot T^{i}_{n}\over 2}
+4\pi \int \bar \epsilon dg +3\bar {\mathcal P}\rbrace
=-12\pi H(r_{0})\bar p(r_{0})g(r_{0})+\int dM C^{i}_{j}
\dot T^{i}_{l}T^{j}_{l}, \eqno 20$$
and from the virial relations we obtain
$$\int dM T^{k}_{l}\ddot T^{i}_{l}=-4\pi\delta^{ik}\int gd\bar p +
3{\mathcal P}^{ik}+\int dM C^{i}_{j}T^{j}_{l}T^{k}_{l}. \eqno 21$$
Differentiating these equations over the mass coordinate we have
$${d\over dt}\lbrace  {\dot T^{i}_{n}\dot T^{i}_{n}\over 2}
+4\pi \bar \epsilon {dg\over dM} -{3\over 2}a^{2}_{j}D_{j}
{GM\over g}\rbrace
=-12\pi{d(H\bar p g)\over dM}+C^{i}_{j}
\dot T^{i}_{l}T^{j}_{l}, \eqno 22$$
and
$$\ddot T^{i}_{n}=-4\pi S^{n}_{i}g{d\bar p\over dM}
-{3\over 2}A^{i}_{j}a_{j}D_{j}E^{j}_{n}{GM\over g}+C^{i}_{j}T^{j}_{n},
\eqno 23$$
The dynamical equations (23) are the main result of this Section.
Note that if the tidal term is absent and the position matrix
is proportional to $\delta^{i}_{n}$: $T^{i}_{n}=a\delta^{i}_{n}$
the equations (23) are reduced to a single equation, which describes
the radial adiabatic oscillations of a star. In that case this
equation follows from the hydrodynamical equations without any approximation.
For the position matrix of a special type $T^{i}_{n}=\hat T^{i}_{n}(t)r_{0}$
our equations are reduced to the dynamical equations of the affine model
(Appendix A).
Also, in the case of an incompressible fluid $\rho(t, x^{i})=\rho_{0}$=const
the equations (23) are exact (see Appendix A).

Eq. (22) must follow from eq. (23). To prove it we contract
both sides of (23) with the velocity matrix $\dot T^{i}_{n}$ over
all indices, and subtract the result from eq. (22). The remainder
can be separated into thermal and gravitational parts. The thermal
part is easily reduced to the relation:
$$d{({\bar \epsilon \over \bar \rho})}+\bar p 
d{({1 \over \bar \rho})}=0, \eqno 24$$
which obviously reflects the application of the
first law of thermodynamics to our case. 
Neglecting the possible presence of shocks, 
for an ideal gas with polytropic index $\gamma $
this equation is integrated to give
$$\bar p =C(r_{0}){\bar \rho}^{\gamma}, \eqno 25$$
where the entropy constant $C(r_{0})$ is determined from an 
unperturbed model of the star. The gravitational part is reduced to the
equality
$${d\over dt}({a_{j}^{2}D_{j}\over g})+{D_{j}a_{j}\dot a_{j}\over g}=0.
\eqno 26$$
Differentiating eq. (19) and using the definition of $D_{j}$,
one can see that in fact, the equality (26) is an identity.

The law of evolution of angular momentum can be easily obtained from (23).
Contracting both parts of (23) with $\dot T^{k}_{n}$ and taking the
antisymmetric part of the result, we have:
$${d\over dt}(T^{k}_{n}\dot T^{i}_{n}-T^{i}_{n}\dot T^{k}_{n})=
C^{i}_{j}T^{j}_{n}T^{k}_{n}-C^{k}_{j}T^{j}_{n}T^{i}_{n}=
a^{2}_{n}A^{j}_{n}(C^{i}_{j}A^{k}_{n}-C^{k}_{j}A^{i}_{n}), \eqno 27$$
where we use eq. (2) to obtain the last equality. Obviously,
the eq. (27) describes the rate of change of angular momentum 
due to a tidal torque. 

Similar to the dynamical system describing the motion of an incompressible
ellipsoid in a tidal field (e. g. Ch, p. 74) and the affine model
(Carter and Luminet, 1985), our system has three additional quantities
which are exactly conserved if the system is evolving in a tidal field
\footnote{Unlike these models, our quantities are not numbers, but
functions of the Lagrangian variable $M$.}.
Contracting (23) with $\dot T^{i}_{m}$, taking the antisymmetric 
part of the result, and using the symmetric properties
of the tidal tensor,
we see that
$$ \chi_{mn}(M)=T^{i}_{n}\dot T^{i}_{m}-T^{i}_{m}\dot T^{i}_{n}, 
\eqno 28$$ 
($\chi_{mn}=-\chi_{nm}$) do not depend on time. 
It is easy to find the physical meaning of $\chi_{mn}$.
For that let us introduce the dual vector 
$\chi_{l}={1\over 2}\epsilon_{lmn}\chi_{mn}$, and consider 
the circulation $C=\oint_{L}v^{i}dx^{i}$, where the integration is
performed over a closed path $L$ on the surface of an ellipsoid of
a given Lagrangian radius. Along this path the vector $e^{i}_{0}$ 
describes a closed curve on a sphere of unit radius. Let us
consider a surface on that sphere enclosed in that curve, and denote 
projections of that surface on the coordinate planes of a Cartesian 
coordinate system where the components $e^{i}_{0}$ are defined,
as $S^{i}$.  Then it is easy to see that
the circulation can be expressed as
$$C=\chi_{i}S^{i}, \eqno 29$$  
and therefore the conservation of the quantities $\chi_{mn}$ corresponds
to the conservation of the circulation of the fluid over our elliptic shells.
It is also interesting to note that the angular momentum tensor and the
quantities $\chi_{mn}$ are adjoint in a certain sense provided the
tidal interaction is switched off. For that it is sufficient to note
that the motion determined by the transpose ${\bf T}^{T}$ of
$\bf {T}$ also provides a solution of the system (23) (for a motion
of an incompressible ellipsoid the similar statement is known 
as Dedekind's theorem). Clearly, 
the quantities $\chi_{mn}$ play the role of angular momentum for the
motion determined by  ${\bf T}^{T}$, and the angular momentum tensor 
plays the role of $\chi_{mn}$. The presence of tidal interactions breaks
this symmetry. 
Since the star is usually assumed to be non-rotating
before the tidal field is "switched on", we set
$\chi_{mn}=0$.

At the end of this Section let us note that similar to
the affine model, the dynamical equations of our model can be deduced
from a variational principle. Consider a Lagrangian of the form:
$${\mathcal L}=\int dM \lbrace {\dot T^{i}_{n}\dot T^{i}_{n}\over 2}
+{3\over 2} a_{j}^{2}D_{j} {GM\over g}+
{C^{i}_{k}T^{i}_{l}T^{k}_{l}\over 2}\rbrace -4\pi\int \bar \epsilon dg, 
\eqno 30 $$ 
The first variance of that Lagrangian can be written as
\footnote{ The variance of the thermal part of (30) is transformed
with help of (5),(24): 
$\delta \int \bar \epsilon dg =\int \delta g d\bar p$.}
$$\delta {\mathcal L}=\int dM \lbrace \dot T^{i}_{n}\delta \dot T^{i}_{n}  
-{3\over 2}A^{i}_{j}a_{j}D_{j}E^{j}_{n}{GM\over g}\delta T^{i}_{n}
+C^{i}_{k}T^{k}_{n}\delta T^{i}_{n} -4\pi g {d\bar p\over dM}S^{n}_{i}
\delta T^{i}_{n} \rbrace \eqno 31$$
Substituting the variance (31) into the Lagrange equations
$${d\over dt}{\delta {\mathcal L}\over \delta \dot T^{i}_{n}}=
{\delta {\mathcal L}\over \delta T^{i}_{n}}, \eqno 32$$
we arrive again at eq. (23). 

\section{Results of numerical calculations}

As we mentioned in the Introduction, for our numerical work  we choose a simple
problem, the tidal encounter of a polytropic star moving around a source
of Newtonian gravity (referred to as a black hole)
on a parabolic orbit. We assume that the star consists of an ideal gas
with constant specific heat ratio $\gamma=5/3$. In this case our problem
can be described by two parameters: a) the polytropic index $n$, and b)
the parameter $\eta $ reflecting the strength of a tidal encounter:
$$\eta=\sqrt{{M_{*}\over M_{h}}{R^{3}_{p}\over R_{*}^{3}}}
={({R_{p}\over R_{T}})}^{{3\over 2}}, \eqno 33$$
where $M_{h}$ is mass of the black hole, $M_{*}$ and $R_{*}$ are the 
mass and radius
of the star, respectively, and $R_{p}$ is the value of the
pericentric separation distance
between the star and the black hole
\footnote{From a physical point of view, $\eta$ is approximately the ratio of
the time which star spends near the pericentric distance to the characteristic 
stellar time $t_{*}=\sqrt {{R_{*}^{3}\over GM_{*}}}$. 
Therefore the strong tidal encounters ($\eta \rightarrow 0$)
are short.}. 
$$ R_{T}= \sqrt [3] {{M_{h}\over M_{*}}}R_{*}\approx
0.91\sqrt [3] {{M_{h}\over \pi \rho_{*}}}, \eqno 34$$ 
is a characteristic ``tidal'' radius, and the average density
$\rho_{*}={3M_{*}\over 4\pi R_{*}^{3}}$. In a very approximate sense it can be said
that 
stars moving on orbits with $R_{p}$ smaller than $R_{T}$ 
(and $\eta < 1$) experience
a strong tidal influence and could be disrupted, and stars moving
on orbits with $R_{p} > R_{T}$ ($\eta > 1$)
are rather weakly perturbed by
the tidal field. In fact, as we see later this conclusion depends on the
polytropic index of the star (see also Kh b)
\footnote{Let us remind that in the classic Roche problem, the infinitesimal
incompressible satellite of density $\rho$ rotating about an object of mass
$M$ in a circular Keplerian orbit, loses its stability if the radius of
the orbit is smaller than $R_{Rh}\approx 2.23\sqrt [3]{ {M\over \pi \rho}}$,
e.g. Ch, p. 12.}.     

We calculate the main characteristics of the tidal disruption event with
a simple explicit conservative
Lagrangian numerical scheme (see Appendix B for the details), and compare
the results of our calculations with the results of 3D simulations reported by Kh a,b.
The values of the polytropic index $n$ are $n=1.5$, $2$, $3$, and
the parameter $\eta $ is changed over a rather broad range. The Eulerian coordinate system
coincides with the Lagrangian coordinate system in the beginning of 
calculations. The plane $(XOY)$ of the Cartesian frame corresponding to the Eulerian
coordinate system coincides with the orbital plane, and the axis $OX$ is directed 
opposite to the black hole when the star passes the point of minimal separation. 
All results of the calculations are expressed in terms of natural units. 
Our spatial unit is the star's radius $R_{*}$, and the time unit is 
$t_{*}=\sqrt {{R_{*}^{3}\over GM_{*}}}$. 
We use dimensionless time $\tau=t/t_{*}$, and the moment of passage of the minimal separation
distance by the star corresponds to $\tau=0$.
The energy and the angular momentum gained
by the star are expressed in units of $GM^{2}_{*}/R_{*}$ and $M_{*}\sqrt{GM_{*}R_{*}}$
respectively. As a Lagrangian coordinate we use the ratio $x$ of mass within a particular
shell to the total mass of the star: $x=M(r_{0})/M_{*}$. 

The results of our calculations are presented in Figures 1-17.

In Figure 1 (a-d) we show projections of our elliptical shells on the plane $(XOY)$ for
four different moments of time $\tau=0$; $1$; $2$; $3$. The shells shown correspond to
four different Lagrangian coordinates $x=0.2$; $0.4$; $0.6$; $0.8$. Although the positions of
the shells have no direct physical meaning since the shells do not describe directly 
the mass distribution over the star and can intersect each other, these
plots give useful qualitative information about the evolution of our model. The model
parameters for these plots $n=1.5$ and $\eta=1.5$ correspond to a tidal encounter of
moderate strength.
Let us recall that one of the principal axes of the tidal tensor is always
oriented toward the black hole, and two others are perpendicular to this direction. Therefore,
as we see from Fig.1 (a-d), the principal axes of the shells do not coincide with the principal
axes of the tidal tensor, and it can be said that our shells lag behind the changing tidal field.
A similar effect has been observed in 3D numerical simulations of the tidal encounter.
At the moment $\tau=0$ all shells are almost aligned with respect to each other, but during the
subsequent evolution this alignment disappears. The outer layers of the star lag more than the 
inner layers, and at the time $\tau=3$ intersections between the shells are observed. Approximately
at the same time in the 3D 
finite difference and SPH models the star takes a typical 
$S$-shaped form.

Figures 2-5 show the evolution of the central density $\rho_{c}$
(expressed in units of the central density
$\rho_{c0}$ of the unperturbed star) with time for the different parameters of the model. The dashed
lines in Figures 2-4 correspond to the results taken from the 3D finite difference simulations
of Kh a,b, and the dashed line in Figure 5 corresponds to the result of SPH simulations by Fulbright
et al 1995. The model parameters for these Figures are: $n=1.5$, $\eta=3$ for Fig. 2a; $n=2$,
$\eta=3$ for Fig. 2b; $n=2$, $\eta=2.5$ for Fig. 2c; $n=2$, $\eta=2$ for Fig. 2d;
$n=3$, $\eta=0.5$ for Fig. 3; $n=2$, $\eta=0.1$ for Fig. 4; $n=1.5$, $\eta=5^{-3/2}\approx 
0.0894$ for Fig. 5. Figs 2(a-d) correspond to rather weak tidal forces; Fig. 3
presents the case of moderately strong tidal disruption and Figs 4,5 present the case
of a very strong tidal disruption event. For the cases of weak tidal influence we observe
oscillations of the star after the moment $\tau=0$. The period of these oscillations is close to
the period $T_{f}$ of the fundamental mode of radial stellar pulsations: $T_{f}=3.81$ for $n=1.5$,
and $T_{f}=\pi$ for $n=2$ (e.g. Cox 1980). It seems that in our model the amplitude of these 
oscillations is always smaller than in the 3D simulations. We checked that this decrease of the
amplitude does not depend on the value of artificial viscosity used in the computations.
Fig. 3 gives an example of a moderately strong tidal disruption event
(only about 10 percent of the mass remains gravitationally bounded after this encounter,
see Kh b and Figure 17). After $\tau=0$
the central density monotonically decreases with time toward a very small number (we obtained
$\rho_{c}/\rho_{c0} \approx 1.4\cdot 10^{-2}$ at the end of our calculations $\tau_{end}=15$).
In general our curve is close to the curve obtained in the 3D simulations, but there is a 
significant deviation near the time $\tau \sim 1$. Unfortunately, the curve obtained
in the 3D case is not continued after $\tau=1$, and we cannot say whether our asymptotic value
for $\rho_{c}/\rho_{c0}$ is close to the 3D result or not. During the tidal disruption events
corresponding to the results presented in Figs 4,5 a very strong tidal disruption of the star
occurs, and a typical 'spike' in the curves for the evolution of the central density is observed.
The presence of this 'spike' has been predicted analytically by Carter $\&$ Luminet $1982$.
The amplitude of this 'spike' is 0.64 times smaller than was obtained by Kh b for the $n=2$, 
$\eta=0.1$ case, and about $1.6$ times larger than was obtained by Fulbright et al 1995
in their SPH simulations of $n=1.5$, $\eta=0.0894$ case. The asymptotic values 
of $\rho_{c}/\rho_{c0}$ at the
limit of large time almost coincide between our curves and the curves obtained by other
methods. 

In Figures 6-9 the dependence of total, potential, thermal and kinetic energies on
time is shown. In all cases we plot the difference between the energy of a particular
kind and its corresponding equilibrium value; $E_{tot}$, $E_{g}$, $E_{th}$ and
$E_{K}$ stand for the differences of total, gravitational, thermal and
kinetic energies, respectively. 
We found that too much energy is stored in the outer layers of
the star in our model, and calculated the energy differences using $97.5$ percent
of the
star's mass
\footnote{Typically we obtained $10-20$ percent 
of the total energy stored in the outer layers of the star at the end of the computation.}. 
The cases shown are: $n=1.5$, $\eta=3$ (Fig. 6); $n=3$, $\eta=1.5$
(Fig. 7); $n=3$, $\eta=1$ (Fig. 8) and $n=3$, $\eta=0.5$ (Fig. 9). Fig. 6 and Fig. 7
correspond to the case of a small tidal influence on the star,
and Figure 8 presents the case of the tidal encounter of a moderate strength. 
As in the density 
plots we see the oscillations of the potential and thermal energies, and these oscillations 
are more pronounced in the 3D simulations (dashed lines). The oscillations of the potential
and the thermal energies seem to compensate each other, and there is no
oscillation of the total and kinetic energies. The relative difference between the
asymptotic
(at large time) values of $E_{tot}$ in our calculation and in the 3D calculations is about
$10$ percent for the $n=1.5$, $\eta=3$ case, $25$ percent
for the $n=3$, $\eta=1.5$ case. The same difference
looks small for the case $n=3$, $\eta=1$ (about $7$ percent), 
but the 3D simulations end at the time
$\tau\approx 3.5$ which is too short to make conclusions about the asymptotic value of the
total energy. Figure 9 presents the case of a strong tidal encounter. The relative difference
of the total energy
$E_{tot}$ is about $20$ percent for that case, but again the 3D simulations end at time
$\tau\approx 1$ and the  asymptotic values of $E_{tot}$ cannot be compared.
We see that $E_{tot}$ and $E_{K}$ continue to grow all the time in that case, but this growth
takes place only for the gas stripped from the star. Let us define the gravitationally bound debris
of the star as a combination of all stars elements where the sum of kinetic and potential
energies is less than zero. The dotted line in Figure 9 presents the difference between
the total energy of the gravitationally bounded debris and the total energy of the unperturbed
star. This difference should tend to an asymptotic value at the large time limit,
which is equal to $0.718$ for that case.

Figures 10-12 show the value of angular momentum in the direction perpendicular to the star's
orbit, gained by the star
during the tidal encounters. The shown cases are: $n=1.5$, $\eta=3$ (Fig. 10);
$n=3$, $\eta=1.5$ (Fig. 11); $n=3$, $\eta=0.5$ (Fig 12). For the cases 
of weak tidal influence the relative deviation between our model and the 3D results
is about $10-20$ percent. For the case of strong tidal disruption the same difference is 
approximately $10$ percent. 
In that case the total angular momentum grows with time, but the angular momentum
of the gravitationally bound debris 
tends to a small asymptotic value (the dotted line in Figure 12).

Now we would like to discuss the dependence of the main characteristics of
a star after
a tidal encounter on the parameter $\eta$ and the polytropic index $n$.
For that we should introduce some quantities describing the perturbed
star which are conserved when the star leaves the region of effective
action of the tidal forces ($\tau \rightarrow \infty $). One 
such quantity is the difference between the total energy 
of the gravitationally bounded debris of the star and the total energy 
of the unperturbed star: $E_{tot}^{bn}$. This quantity sharply decreases
with increase of the parameter $\eta$, and therefore it is convenient
to introduce another quantity $T(\eta)=\eta^{4}E_{tot}^{bn}$. $T(\eta)$
can be easily compared with results of linear theory of tidal 
perturbations of a star (Press $\&$ Teukolsky 1977, 
Lee $\&$ Ostriker 1986), and with results of 3D simulations. 
We show the result of calculations of this quantity in Figures 13
for a polytropic star with $n=3$, in Figure 14 for the case $n=2$, and  
in Figure 15 for the case $n=1.5$, respectively
\footnote{ For the calculations reported hereafter we used the time
interval $-10 < \tau < 10$.}.
One can see that the difference between the linear theory and our model
is small if $\eta $ is rather large ($\eta \sim 2$ for the $n=3$ case, 
and $\sim 3.5-4$ for the $n=2$, $1.5$ cases),
but the linear theory underestimates significantly the energy gain
at intermediate values of $\eta$ ($\eta \sim 1, \sim 1.5, \sim 2$ 
for the $n=3$, $2$, $1.5$ cases respectively). 
This fact is confirmed by the 3D simulations (open circles in Figures
13-15), and is very transparent from a qualitative point 
of view. Indeed, the action of tidal forces on the star causes an increase
of a characteristic size of the star. In turn, this increase leads to
an increase of the characteristic amplitude of the tidal forces.
Since the $n=3$ polytrope is
more concentrated toward the stellar center of mass
than the $n=2$ case, and the $n=2$ polytrope is more concentrated
than the $n=1.5$ case, for a given $\eta$ the energy gain of 
the $n=3$ polytrope is always smaller than the energy gain of the $n=2$,
and the energy gain of 
the $n=2$ polytrope is respectively 
smaller than the energy gain of the $n=1.5$ case.
Note, that the results corresponding to large values of 
$\eta$ should be taken with care. At first our model may not
give exactly the same results as the linear theory in the limit
of small tidal perturbations.  Second in our numerical model the difference 
between the numerical value of the total energy for an unperturbed star
and its theoretical value is of the order of $10^{-2}$, and this
difference is larger than the energy gain
corresponding to  large values of $\eta$ (e.g $\eta \geq 2$ for the $n=3$ case). 
Also the circle corresponding
to $\eta=0.5$ is shown for illustrative purposes only. For those values of
$\eta$ a significant stripping of the star occurs (see Figure 17), 
and a deviation of the total energy of the star (shown by the circle) and the
total energy of the gravitationally bound debris can be significant.
In Figure 14 the open circles almost coincide with our curve except for
the circle
corresponding to $\eta=1.5$, where the difference is rather large (about
40 percent). Perhaps this difference is related to effect of the  partial
mass stripping of the star, which is rather pronounced for that case.
This effect makes the determination of energy of the
gravitationally bound debris
rather ambiguous, 
since the gas leaves the star with almost parabolic velocities
for this tidal encounter of a moderate strength. 

In Figures 16 we show the values of the angular momentum perpendicular to
the star's orbit (taken at the end of calculations $\tau_{end}=10$) and 
for the gravitationally bound debris
(the dotted line) versus $\eta$ for the $n=3$ polytrope.
The open circles correspond to the 3D simulations and the dashed curve
represents the same quantity calculated by Diener et al
1995 in the framework of the affine model. It seems that in our model the star
gets more angular momentum than in the affine model. 

In general Figures 13-16 show very good agreement between our results and the results
of the 3D simulations. Unfortunately the number of the 3D experiments is too
small to make a robust quantitative comparison. 

In Figure 17 we show the amount of mass lost by the star after a fly-by around
the black hole as a function of $\eta$. 
Contrary to a 'naive' Roche-like criterion of tidal disruption (the
discussion after eq. 34) we see that there is a partial stripping of mass
from the star for a particular range of $\eta$. For the
$n=3$ polytrope (the solid line) the star loses 
its mass if $\eta < \eta_{strip}\approx 1.5$,
and the star is completely disrupted if $\eta < \eta_{crit}\approx 0.4$.
For the $n=2$ polytrope (the dashed line) we have 
$\eta_{strip}\approx 2$ and $\eta_{crit}\approx 0.9$ and
for the $n=1.5$ polytrope (the dotted line) we have 
$\eta_{strip}\approx 2.5$ and $\eta_{crit}\approx 1.14$. The values of 
$\eta_{strip}$
and $\eta_{crit}$ are in excellent agreement with estimates based on the
3D simulations (see Kh b). It is instructive to compare these results with a criterion
of tidal disruption obtained in the affine model. In this model Diener et al 
found that the star is disrupted if $\eta < \eta^{a}_{crit}=0.844$ for $n=3$,
$\eta < \eta^{a}_{crit}=1.482$ for $n=2$, 
$\eta < \eta^{a}_{crit}=1.839$ for $n=1.5$.  Since in the affine model the velocity
field is self-similar, in this model  
a partial stripping of outer layers of the star
cannot occur. 
It can be observed that the inequalities $\eta_{crit} < \eta^{a}_{crit}
< \eta_{strip}$ always hold, and the difference between
the values of $\eta_{crit}^{a}$ and
values $\eta_{crit}$ is about 40-50 percent. This difference is much larger than the
difference between our values and the results of the 3D calculations. This demonstrates
the advantage of our model.

\section{Conclusions and Discussion}

We have constructed a new model of a star perturbed by a tidal field. In this model the 
star consists of
a set of elliptical shells which in general 
have different principal axes and different orientations with respect to a fixed locally inertial
reference frame comoving with the star's center of mass. The model obeys certain 
evolution equations. The results of calculations of a simple problem
of tidal encounter of a polytropic star moving on a parabolic orbit around a source of Newtonian
gravity have been compared with the results of three dimensional finite difference simulations.
We found that the main characteristics of the tidal encounter agree with the results
of the finite difference approach with a
typical accuracy $10-20$ percent.
Taking into account that the astrophysical applications do not demand 
very high accuracy in description of a single tidal encounter, 
we think that our model could be used in order to investigate
all possible variants of the problem of tidal interaction between a supermassive black hole and
a star interesting from an astrophysical point of view.
The main advantage of our model is its effectively one dimensional character, which allows us
to calculate all interesting variables much faster than the 3D approach, and over a much
longer time of evolution. Also, all characteristics of the tidally perturbed star could be
inferred from our model in a straightforward and unambiguous way. Our model could also be used 
in a study of a rotating single star, or a binary star.

In principal, the agreement between our model and 3D computations could be improved if one
uses more advanced variants of our model (see below). However we would like to note that in
order to make a comparison between  an advanced variant of our model and 3D computations,
one should also increase both the number of the numerical experiments and their resolution.
For example, recently it has been claimed (Ayal et al, 2000)
that the difference between the different numerical experiments in SPH models is about 10 
percent depending on number of particles used in the calculations. We are not aware of 
similar convergence studies for the 3D finite difference models. We think that such studies should
be undertaken parallel to work on improvement of our model.  

Now we would like to discuss possible extensions of our formalism. One obvious
way for such extension consists in using a more complicated stellar model for the unperturbed state.
Then the realistic stellar models could be generalized to our problem by using 
the energy evolution equation and the virial relations similar to the eq. (8,9), 
but written for a realistic stellar gas, with possible inclusion of e.g. non-adiabatic effects,
viscosity, effects of radiative transfer, and generation of heat due to nuclear reactions 
in the stellar core. The evolution equations for such a model could be derived
from the energy equation and the virial relations in the way described above. One could also
use a more refined numerical scheme, say an implicit scheme with a nonuniform grid. 
We suppose that certain powerful methods 
developed in numerical investigations of pulsating stars could be
directly applied to our problem. Another interesting extension consists in using the real 
distribution of the pressure and the density over the volume of the star in our approximation.
Thus it is possible to construct a model with no intersection between the shells,
which could provide more information about displacements of particular elements of the
star during the tidal encounter. Assuming that the star consists of an ideal gas 
with constant ratio of specific heats $\gamma$, and neglecting the possible 
presence of shocks in the system, this problem is reduced to the evaluation of the following
integrals:
$$I_{1}=\int_{S}R^{2(\gamma -1)}d\Omega, \eqno 35$$
and
$$I_{2i}=\int_{S}l^{2}_{(i)}R^{2\gamma}d\Omega, \eqno 36$$
where the integration is performed over the surface of an ellipsoid
$\sum \lambda_{(i)}y^{2}_{(i)}=1$. $\lambda_{i} > 0$ are the eigenvalues
of the matrix $R^{ln}$ (see. eq. 4), $R$ is the value of the radius vector joining
the center of the ellipsoid to a point on its surface, $l_{i}$ are the direction
cosines of the radius vector, and $d\Omega$ is the elementary solid angle. The
Cartesian coordinates $y_{i}$ are associated with the frame of eigenvectors of
the matrix $R^{ln}$. If $\gamma=1$ the integral (35) is trivial, and the integrals 
(36) are related to the integrals $D_{i}$ defined above (eq. 18), 
but with $\lambda_{j}^{-1}$
playing the role of $a_{j}^{2}$.  One can 
see that the pressure tensor (defined as the left hand side of the eq. 13)
can be evaluated with help of the integrals (35, 36). The volume part of the
pressure tensor (the first term on l.h.s. of the eq. (13)) is obtained by integration over
the mass of a quantity proportional to the integral (35) with the upper limit
of integration determined by some given Lagrangian coordinate $x_{*}$. The surface term ( the second term
on l.h.s. of the eq. (13)) is expressed with help of the integrals (36) with
$\lambda_{i}=\lambda_{i}(x_{*})$. Note that now the surface term is not symmetric, and therefore
transfer of angular momentum between the neighboring shells due to pressure is allowed.
If some shells are close to intersection, the density at some
particular value of $x$ tends to infinity. That means that some eigenvalues $\lambda_{i}$ go to zero,
and as a consequence the integrals (35), (36) tend to infinity causing an increase of pressure.
In turn the increase of pressure could prevent the shells from intersecting.
The integrals (35), (36) can be evaluated e.g. by numerical means, 
and serve as main building blocks of a model
without intersections between the shells corresponding to different Lagrangian radii.

Finally one can generalize our model to the case of a relativistic tidal field. In a separate paper
we apply our model to the problem of tidal interaction of 
a star with a supermassive Kerr black hole.

\acknowledgments

We are grateful to P. Diener, I. Igumenshchev, A. Khokhlov, O. Rognvaldsson, and V.
Turchaninov for useful remarks concerning the numerical methods, to J. Schmalzing for
very useful comments, to
E. Kotok for the help in preparation of this paper,
and also to the referee for very helpful suggestions. 
It is our sincere pleasure to
thank A. Doroshkevich and V. Frolov for many stimulating and fruitful discussions.
This work was supported in part by RFBR grant N 00-02-161335 and
in part by the Danish Research Foundation through its
establishment of the Theoretical Astrophysics Center.



\appendix 

\section{Reduction of the dynamical equations to the equations of the affine model
and the case of an incompressible fluid}

Let us consider the position matrix of a special form:
$${\bf T}(t, r_{0})=\hat {\bf T}(t)r_{0}, \eqno A1$$
and introduce the matrices $\hat {\bf A}(t)$,  $\hat {\bf E}(t)$,  $\hat {\bf S}(t)$,
and the quantities $\hat a_{i}(t)$ defined with the help of
the matrix $\hat {\bf T}(t)$ by analogy with the matrices
${\bf A}$, ${\bf E}$, ${\bf S}$, and the quantities $a_{i}$, respectively.
For the position matrix of the form (A1) the matrix $R^{ln}$ is:
$$R^{ln}={\delta^{ln}\over r_{0}}, \eqno A2$$
and the Jacobian $D=\hat g(t) \equiv \hat a_{1} \hat a_{2} \hat a_{3}$ does not depend on
the spatial coordinate.
Therefore in this case there is no difference between the exact and the averaged
values of the density and the pressure. Substituting the expression (A1) into the
equations (23), multiplying the result on $r_{0}$ and integrating over $M$, we obtain
$$\ddot {\hat T^{i}_{n}}={\Pi \over M_{*}}\hat S^{n}_{i}+{\Omega_{*}\over 2M_{*}}
{\hat g}^{-1}\hat A^{i}_{j}\hat a_{j}\hat D_{j}
\hat E^{j}_{n} +C^{i}_{j}\hat T^{j}_{n}, \eqno A3$$
where $\Pi=\int dM {P\over \rho}$, the constant factor $M_{*}={1\over 3}\int dM r_{0}^{2}$ is
the scalar quadrupole moment of the star in the unperturbed state
and the constant factor $\Omega_{*}=-G\int{MdM\over r_{0}}$ is 
the self-gravitational energy value in the unperturbed state. The quantities 
$\tilde D_{j}$ are defined with the help of the quantities $\tilde a_{j}$ by analogy with 
the quantities $D_{j}$. The equations (A2) are just the dynamical 
equations of the affine model (e.g. Carter $\&$ Luminet 1983, 1985 and Luminet $\&$ Carter,
1986). 
  
Now let us consider the case of an incompressible fluid. In this case the density $\rho$ is
not changing durind the motion and therefore the constraint $\hat g=1$ must be imposed, see
eq (6).
Assuming that the density is also constant over the whole volume of the star 
$\rho=\rho_{0}$=const,
the equations (A3) can be reduced to a very simple form. Taking into account the constraint
$\hat g=1$, and performing the integration over $M$ in the expressions for $\Pi$,
$M_{*}$, and $\Omega_{*}$, we have
$$\ddot {\hat T^{i}_{n}}={\hat S^{n}_{i} P_{c}(t)\over 2 \rho_{0} R_{*}^{2}}-
2\pi G\rho_{0} \hat A^{i}_{j}\hat a_{j} \hat D_{j} \hat E^{j}_{n}+
C^{i}_{j}\hat T^{j}_{n}, \eqno A4$$
where $R_{*}$ is the radius of the star and $P_{c}(t)$ is the pressure in 
the center of the star.
Together with the constraint $\hat g =1$ the equations (A4) form a complete set of
equations.
These equations have
already been obtained by Luminet and Carter (see eq. (3.1a) in Luminet $\&$ Carter, 1986)
\footnote{Note that there is a misprint in the equation (3.1a). The factor 
${1\over 2\rho_{0} R_{*}^{2}}$ in the last term on the right hand side is missed.}.
It has been proved that the equations (A4) follow from the hydrodynamical equations
without any approximation.

\section{The numerical scheme}

For our numerical work it is convenient to introduce the
potential energy tensor density per unit of mass:
$$F^{ik}={d\over dM}\bar {\cal P}^{ik}=-{1\over 2}GMA^{i}_{l}A^{k}_{l}{a^{2}_{l}D_{l}
\over g}, \eqno B1$$
and use dimensionless representations of all variables. We use the
dimensionless mass coordinate $x=M/ M_{*}$, and the dimensionless time 
$\tau=\sqrt{{GM_{*}\over R_{*}^{3}}}t$. $\Delta x$ stands for the mass
coordinate step, and $\Delta \tau$ stands for the time step. The dimensionless dynamical 
variables are defined as follows:
$$\tilde {\bf T}={\bf T}/R_{*}, \quad \tilde {\bf V}=
{\partial \over \partial \tau}\tilde {\bf T}, \eqno B2$$
$$\tilde {\bf F}={R_{*}\over GM_{*}}{\bf F}, \eqno B3$$
$$\tilde g = g/R_{*}^{3}, \quad \tilde \rho ={4\pi R_{*}^{3}\over 3M_{*}}\rho, \eqno B4$$
$$\tilde P ={4\pi R^{4}_{*}\over GM_{*}^{2}}P, \quad 
\tilde \epsilon ={\tilde P \over (\gamma -1)}, \eqno B5$$
the dimensionless tidal tensor is
$$\tilde {\bf C}={R_{*}^{3}\over GM_{*}} \bf C. \eqno B6$$ 

We use a uniform grid along the mass coordinate with number of the grid points $J=201$
(the first and the last numbers
correspond respectively to the center of the star, and to the star's
boundary). The pressure, the density and the energy density are defined
at the centers of zones between the grid points, and are denoted by the indices
$j+1/2$, $j-1/2$, and the position matrix and its time derivative are defined
at the grid points. 
Hereafter the index $n$ denotes a time level, and the index $j$ denotes
a particular grid point, and no summation over those indices is assumed.  

We use several intermediate steps to advance our model from $\tau^{n}$ to $\tau^{n+1}$.
First we calculate $\tilde {\bf C}$, and
$\tilde g^{n}_{j}$ and $\tilde {\bf F}^{n}_{j}$ with help of special subroutines. 
To calculate the rotational matrix ${\bf A}$ and the principal axes $a_{l}$ we use the
standard Jacobi method.  Then we calculate the dimensionless expansion rate
$$\tilde H^{n}_{j}={1\over 3}(\tilde {\bf S} \cdot \tilde {\bf T})^{n}_{j}, \eqno B7$$ 
where $\cdot $ stands for the contraction over all indices, and the density 
$$\tilde \rho^{n}_{j-1/2}={\Delta x\over (\tilde g^{n}_{j} -\tilde g^{n}_{j-1})},
\eqno B8$$ 
Next the artificial
(bulk) viscosity term is calculated
according to the rule:
$$v^{n}_{j}=(\tilde H {\tilde g}^{1/3})^{n}_{j}, \quad
q^{n}_{j-1/2}=b^{2}\tilde \rho^{n}_{j-1/2}{(\Delta x)}^{2}{(v^{n}_{j}-v^{n}_{j-1})}^{2},
\eqno B9$$
if $v^{n}_{j}-v^{n}_{j-1} < 0$, and
$$q^{n}_{j-1/2}=0,$$
if $v^{n}_{j}-v^{n}_{j-1} > 0$.
The artificial viscosity is very useful in the calculations, since it dampen the small scale
(unphysical) modes. We used $b=2$ in our calculations, and 
checked that smaller values of $b$ lead to the same motion, but with additional
small scale oscillations of the dynamical variables. 
After calculation of all these variables we advance the position matrix and its
time derivative to the next time level:
$$\tilde {\bf v}^{n+1}_{j}=\tilde {\bf v}^{n}_{j}+\Delta \tau (-(\tilde {\bf S}^{T}\tilde g)^{n}_{j}
{\partial \tilde P \over \partial x}+3{(\tilde {\bf S}^{T}\tilde {\bf F})}^{n}_{j}
+{(\tilde {\bf C} \tilde {\bf T})}^{n}_{j}), \eqno B10$$
where
$${\partial \tilde P \over \partial x}=
{(\tilde P^{n}_{j+1/2}+q^{n}_{j+1/2}-\tilde P^{n}_{j-1/2}-q^{n}_{j-1/2})\over \Delta x},$$   
and
$$\tilde T^{n+1}_{j}=\tilde T^{n}_{j}+\Delta \tau \tilde {\bf v}^{n+1}_{j}. \eqno B11$$
Note, that the scheme (B10-B11) conserves the circulation, and it also conserves the angular
momentum provided the tidal term is 'switched off'. To calculate the change of the energy density
and the pressure, we calculate the kinetic energy term ${\dot {\bf T}\cdot\dot {\bf T}^{T}
\over 2}$, and the potential energy term $ tr({\bf F})$ for the both time levels, and also the
density $\tilde \rho^{n}_{j-1/2}$ for the $n+1$ time level. Then we have
$$\tilde \epsilon^{n+1}_{j-1/2}=\tilde \rho^{n+1}_{j-1/2}({\tilde
\epsilon^{n}_{j-1/2}\over \tilde \rho^{n}_{j-1/2}}
-(\Delta E_{k} +\Delta E_{p})+\Delta \tau Q), \eqno B12$$ 
where
$$\Delta E_{k}={({\dot {\bf T} \cdot \dot {\bf T}^{T}\over 2})}^{n+1}_{j}-{({\dot {\bf T}
\cdot \dot {\bf T}^{T}\over 2})}^{n}_{j} \quad
\Delta E_{p}= 3tr({\bf F}^{n+1}_{j}-{\bf F}^{n}_{j}), \eqno B13$$
and
$$Q=-3{(\tilde H^{n}_{j}\tilde g^{n}_{j}(\tilde P^{n}_{j+1/2}+q^{n}_{j+1/2})-
\tilde H^{n}_{j-1}\tilde g^{n}_{j-1}(\tilde P^{n}_{j-1/2}+q^{n}_{j-1/2}))\over \Delta x}
+{(\tilde {\bf C} \tilde {\bf T}\cdot \tilde {\bf V}^{T})}^{n}_{j}. \eqno B14$$
The eq. (B12) tells that the scheme is conservative with respect to the law of energy 
conservation. 

The boundary conditions for our problem are
$${\bf T}^{n}_{1}={\bf V}^{n}_{1}=0, \quad \tilde \rho^{n}_{1/2}=\tilde \rho^{n}_{3/2}, \eqno B15$$
and
$$\tilde P^{n}_{J}=0. \eqno B16$$
The initial static stellar configurations are calculated by a shooting method described by e.g.
Khokhlov 1991. 

The size of the time step must follow from the stability analysis of our scheme, 
which is rather complicated. Therefore we constrain our time step by the condition:
$$\Delta \tau ={\alpha \Delta x \over c_{s}}, \eqno B17$$
where
$$c_{s}={\Delta x\over {\tilde g\sqrt { \gamma (\tilde P + q)\tilde \rho (\tilde {\bf S} \cdot
\tilde {\bf S}^{T})}}}, \eqno B18$$
is the velocity of propagation of a small perturbation (with respect to the mass coordinate)
calculated in analytical linear approximation,
and $\alpha$ is a parameter. We used $\alpha =1/15$ for all our calculations except the case
$n=2$, $\eta=0.1$, where $\alpha=1/60$ has been used, and the case 
$n=1.5$, $\eta=0.0894$, where we used $\alpha=1/120$. In the case of the evolution of a pulsating
star our condition (B17) is reduced to the well known von Neumann-Richtmyer condition provided
$\alpha=1/\sqrt{(4\pi)}$ (e.g Richtmyer $\&$ Morton, 1967, p. 297-298). 

We tested our scheme against Sedov's analytical solution for a motion of an expanding 
spherical gas cloud,
and also the problem of small oscillations in a polytropic star. In the both cases we found very good
agreement between the analytical theory and our scheme.





\clearpage



\begin{figure}
\plotone{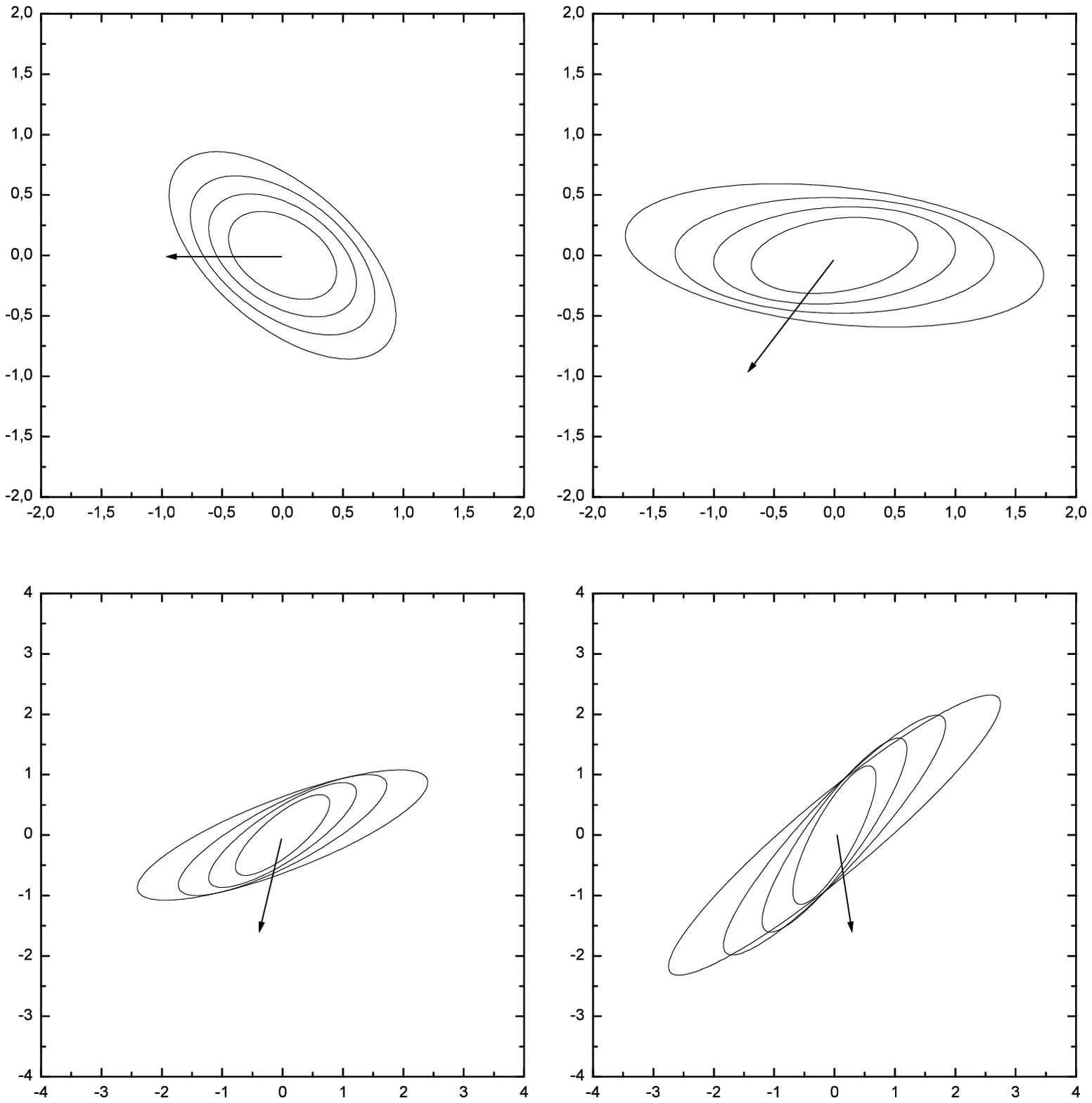}{}
\vspace{-6cm}
\caption
{a) (upper left) The projections of the elliptical shells on the plane $(XOY)$ are shown.
The origin of the coordinate system is centered at the star's center of mass, and the plane
$(XOY)$ lies in the orbital plane. The spatial scales are expressed in units of the stellar 
radius $R_{*}$. 
The ellipses shown correspond to four different Lagrangian
coordinates $x=0.2$ (the innermost ellipse), $x=0.4$, $x=0.6$, $x=0.8$.
The time $\tau=0$ corresponds to
the time of passage of pericenter of the orbit
by the star. The arrow is directed toward the hole.  
b) (upper right) The same as Fig. 1a, but 
$\tau=1$.
c) (bottom left) The same as Fig. 1a, but 
$\tau=2$.
d) (bottom right) The same as Fig. 1a, but 
$\tau=3$.
\label{fig1a,b,c,d}}
\end{figure}

\begin{figure}
\plottwo{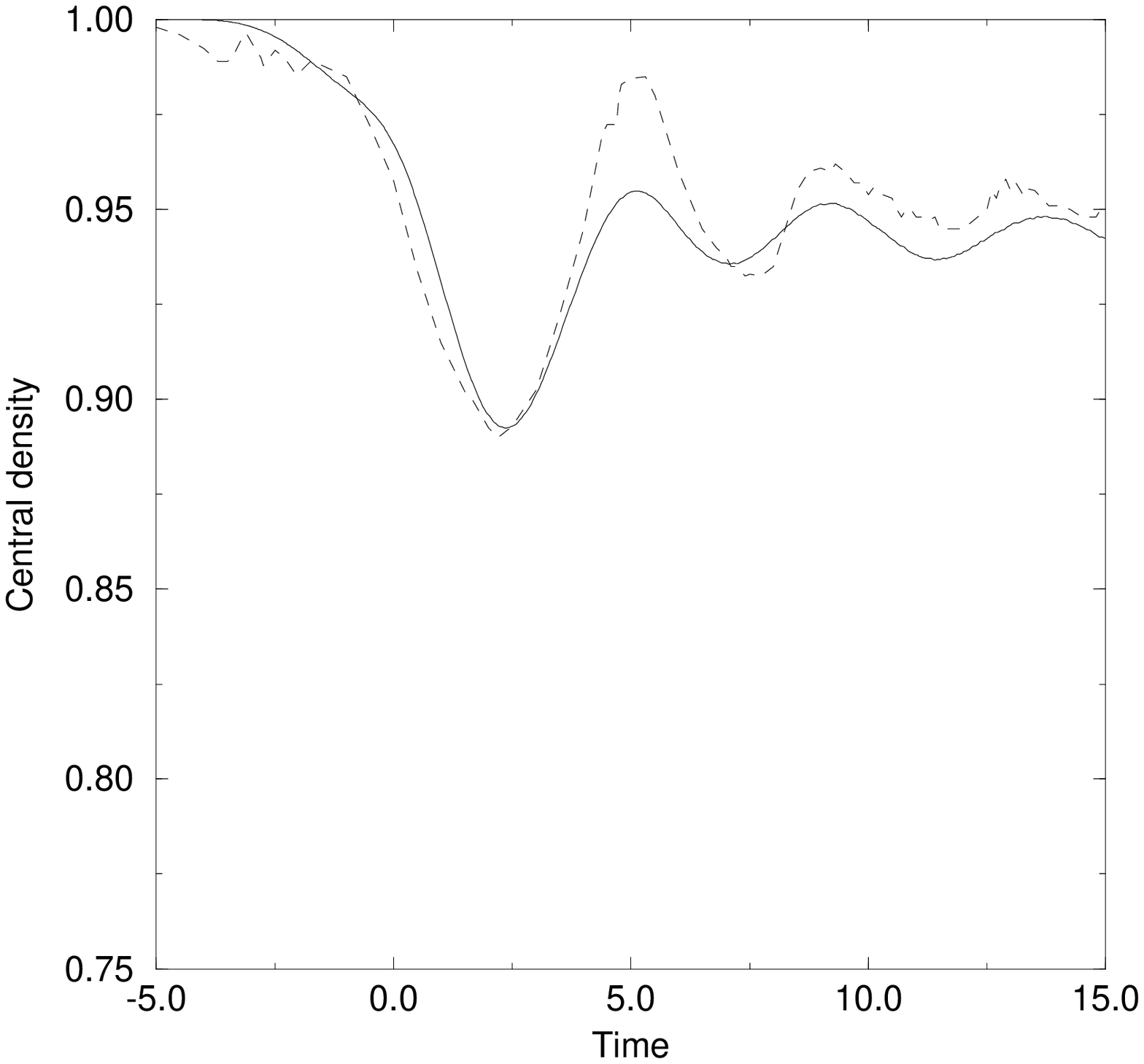}{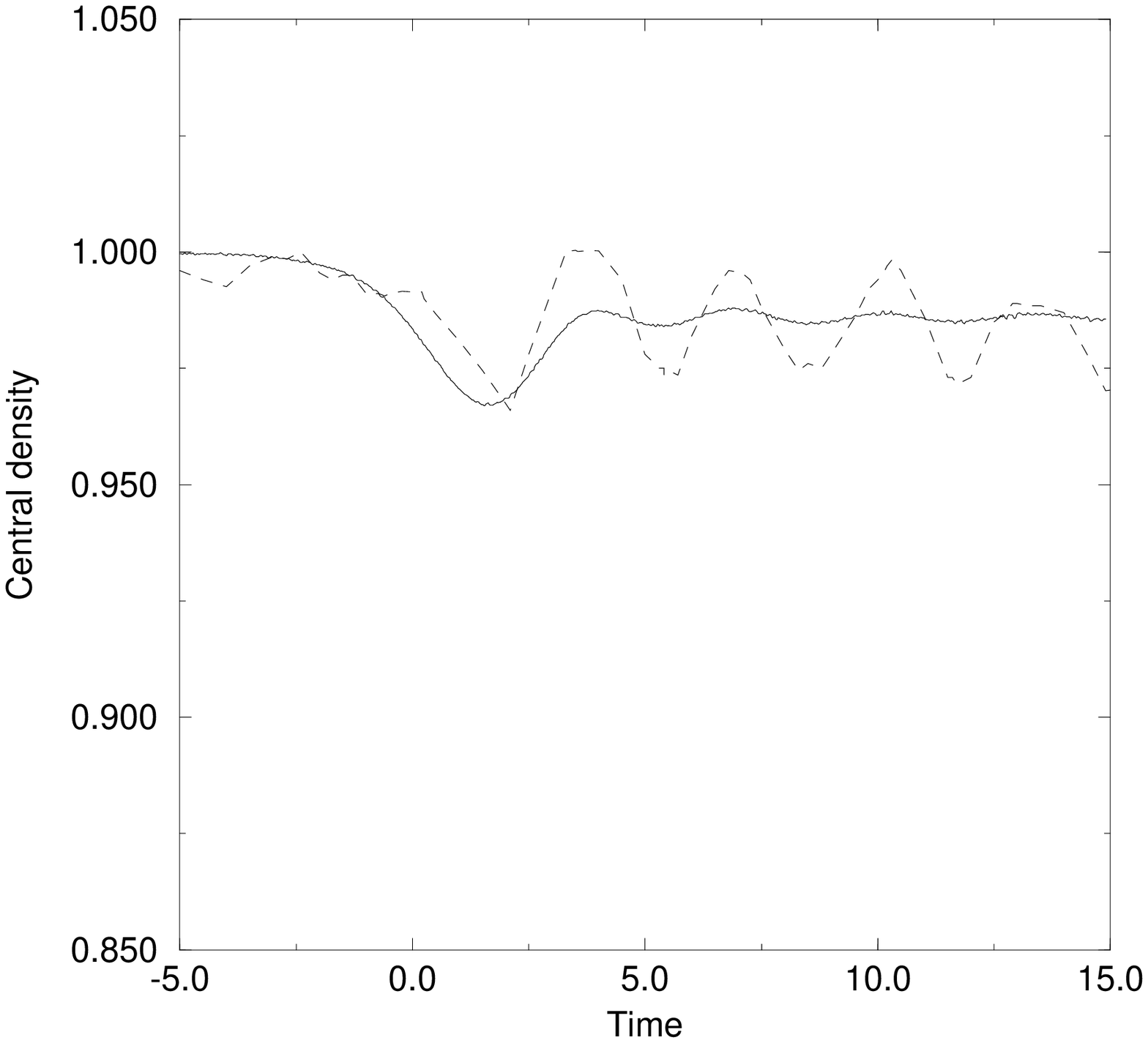}{}
\plottwo{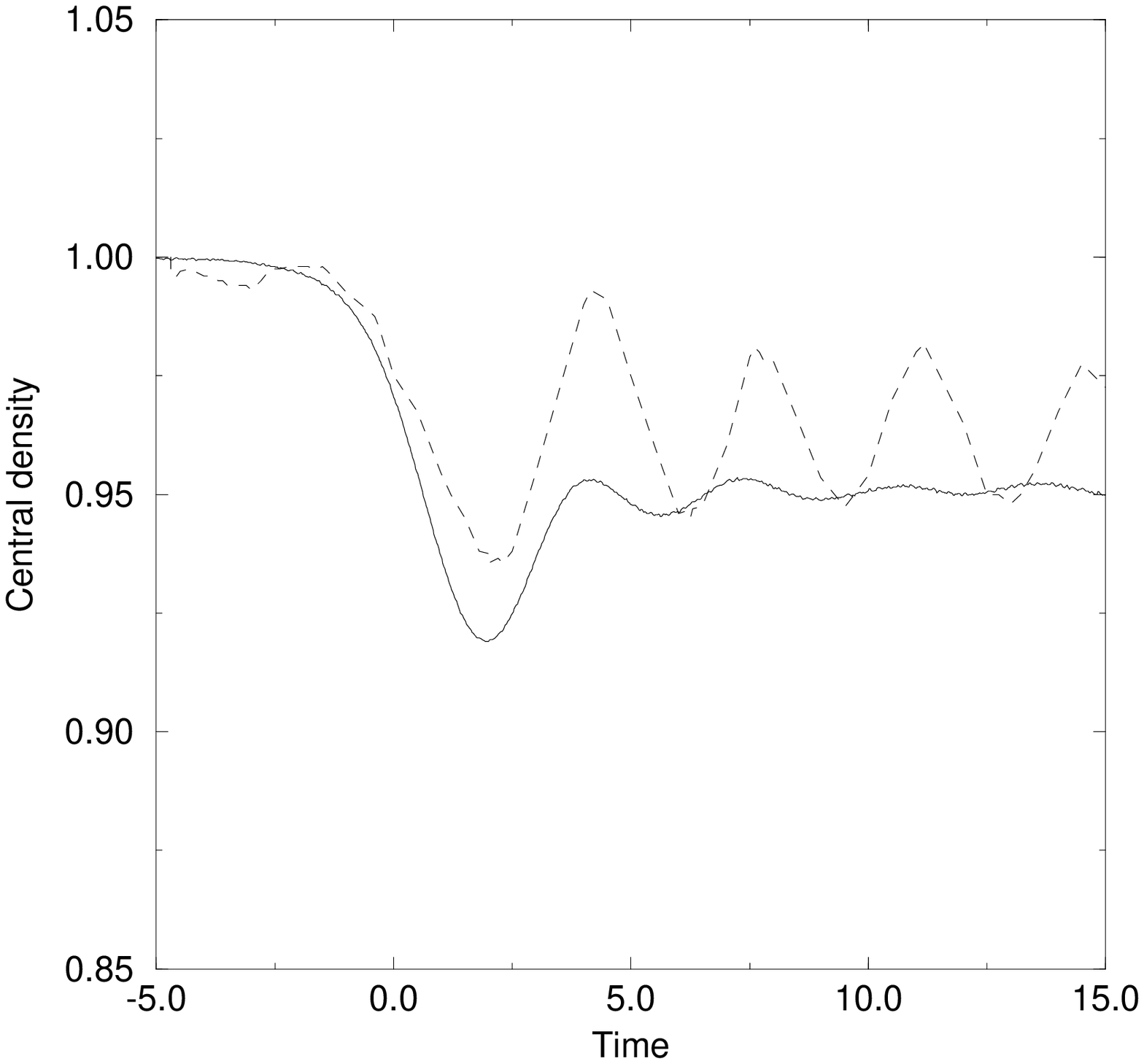}{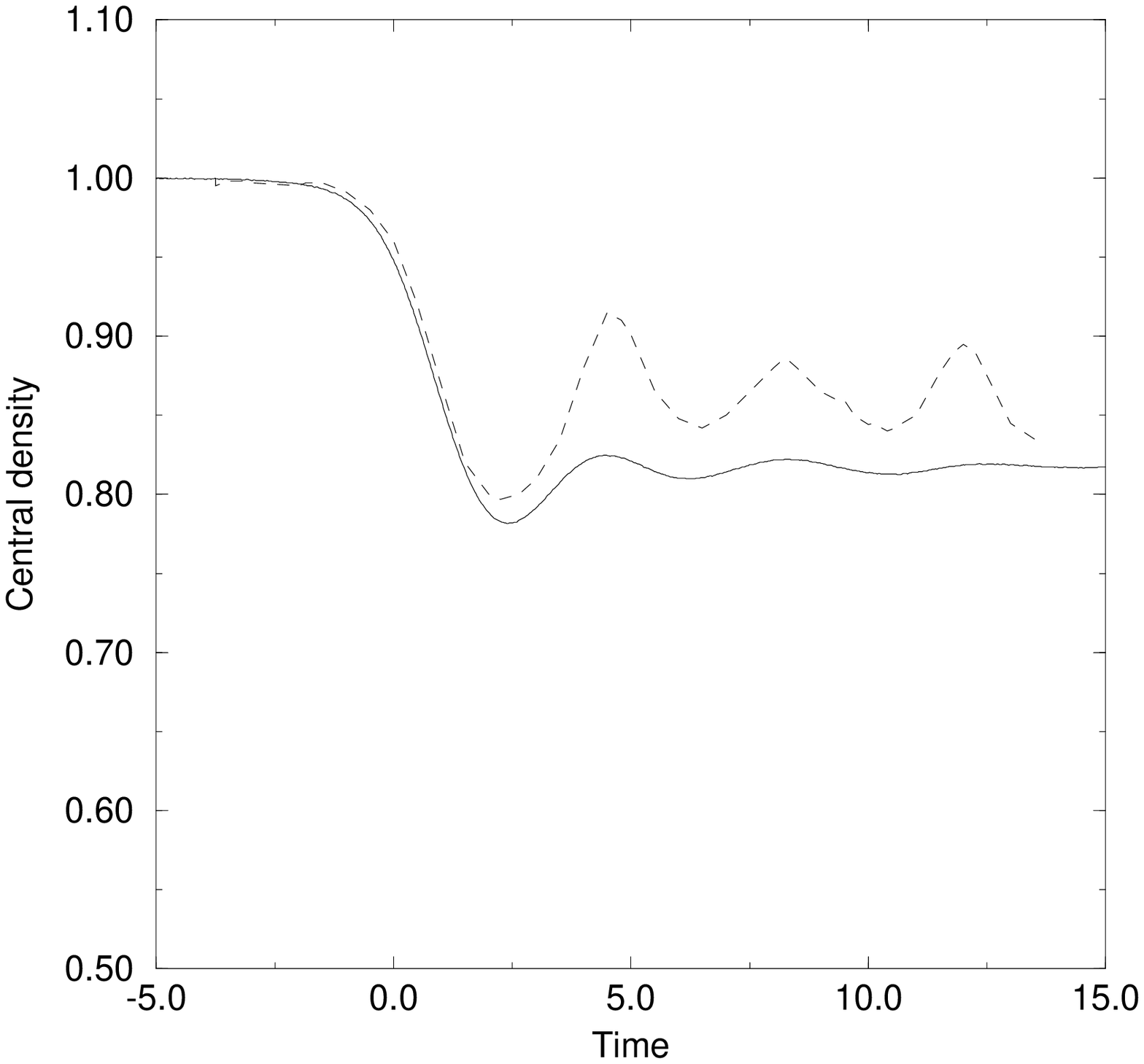}{}
\caption
{a) (upper left) The evolution of the 
central density (expressed in units of the central density of
the unperturbed star) 
is shown as a function of time. The parameters of tidal encounter are
$n=1.5$, $\eta=3$. 
The solid curve is calculated from our model, and the dashed curve is calculated in
3D simulations of Kh a. b) (upper right)
The same as Fig. 2a, but $n=2$, $\eta=3$. c) (bottom left) 
The same as Fig. 2a, but $n=2$, $\eta=2.5$. d) (bottom right) 
The same as Fig. 2a, but $n=2$, $\eta=2$.
\label{fig2a,b,c,d} }
\end{figure}

\begin{figure}
\plotone{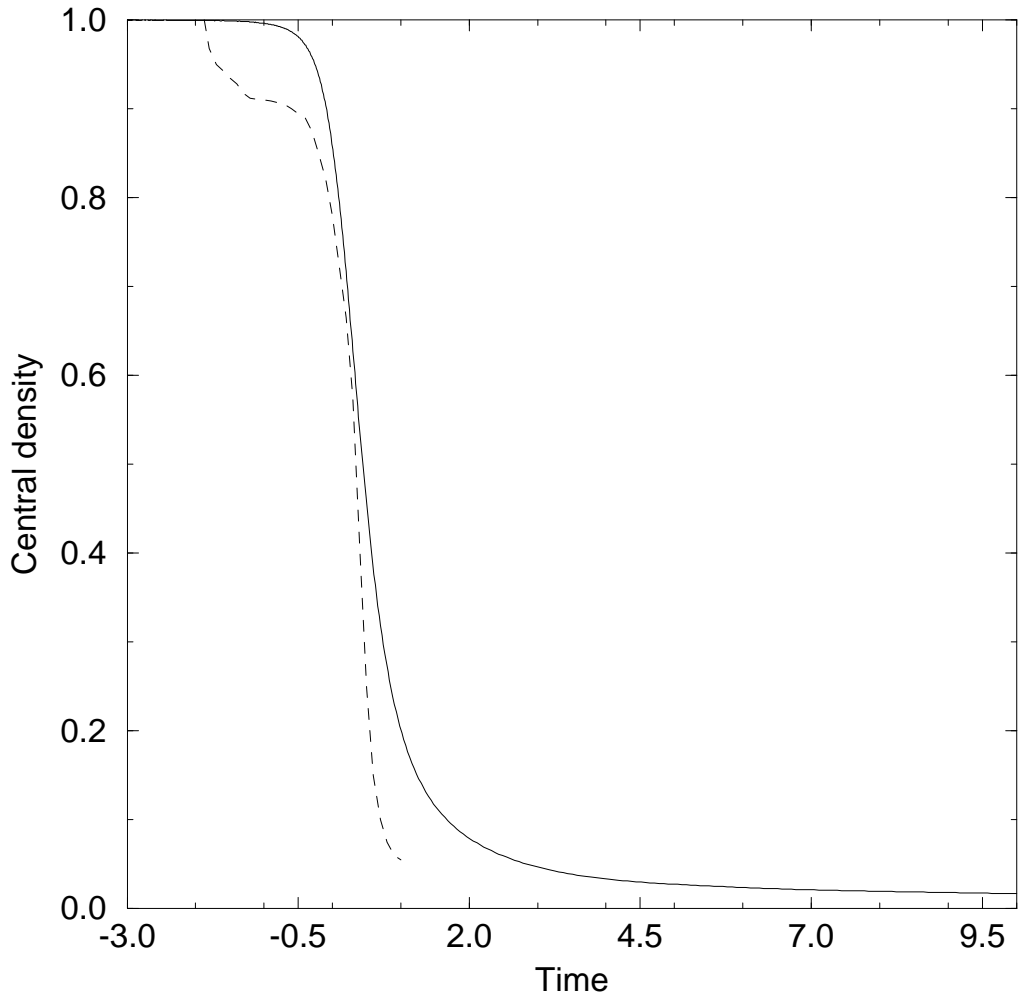}
\caption
{The evolution of the central density in case of a strong
tidal encounter. The dashed curve corresponds to the simulations of Kh b. 
The parameters are $n=3$, $\eta=0.5$.
\label{fig3}}
\end{figure}

\begin{figure}
\plotone{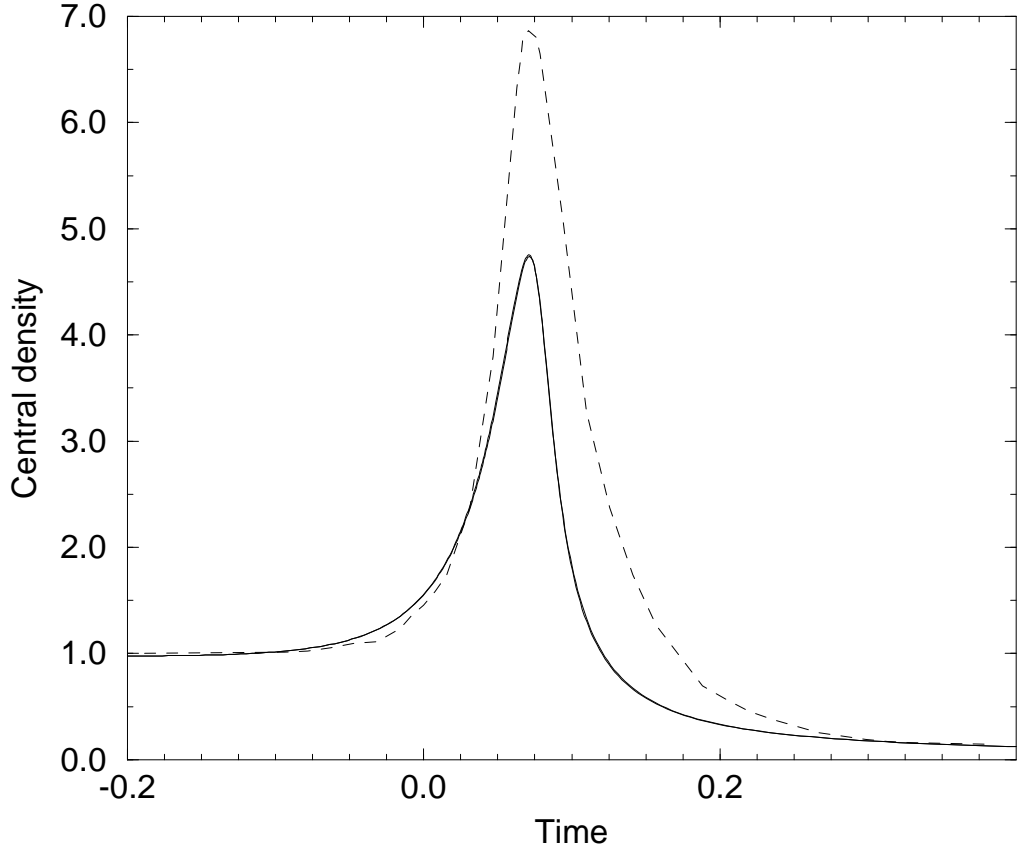}
\caption
{The 'spike' in the evolution curve of the central density due to
very strong tidal encounter, the dashed curve corresponds to the simulations of Kh b. 
The parameters are $n=3$, $\eta=0.1$.
\label{fig4}}
\end{figure}
\begin{figure}
\plotone{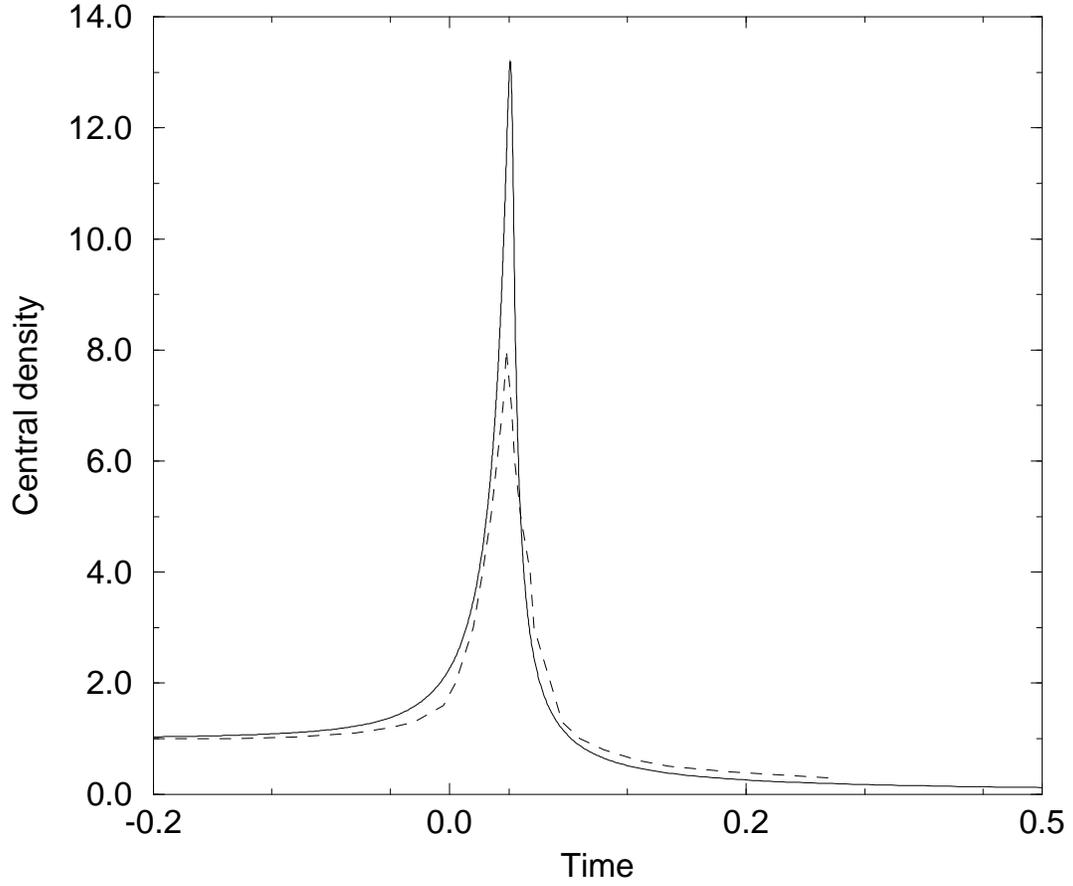}
\caption
{The same as Fig. 4, but $n=1.5$, $\eta=0.0894$. The dashed curve is calculated
in SPH simulations of Fulbright et al, 1995.
\label{fig5}}
\end{figure}

\begin{figure}
\plotone{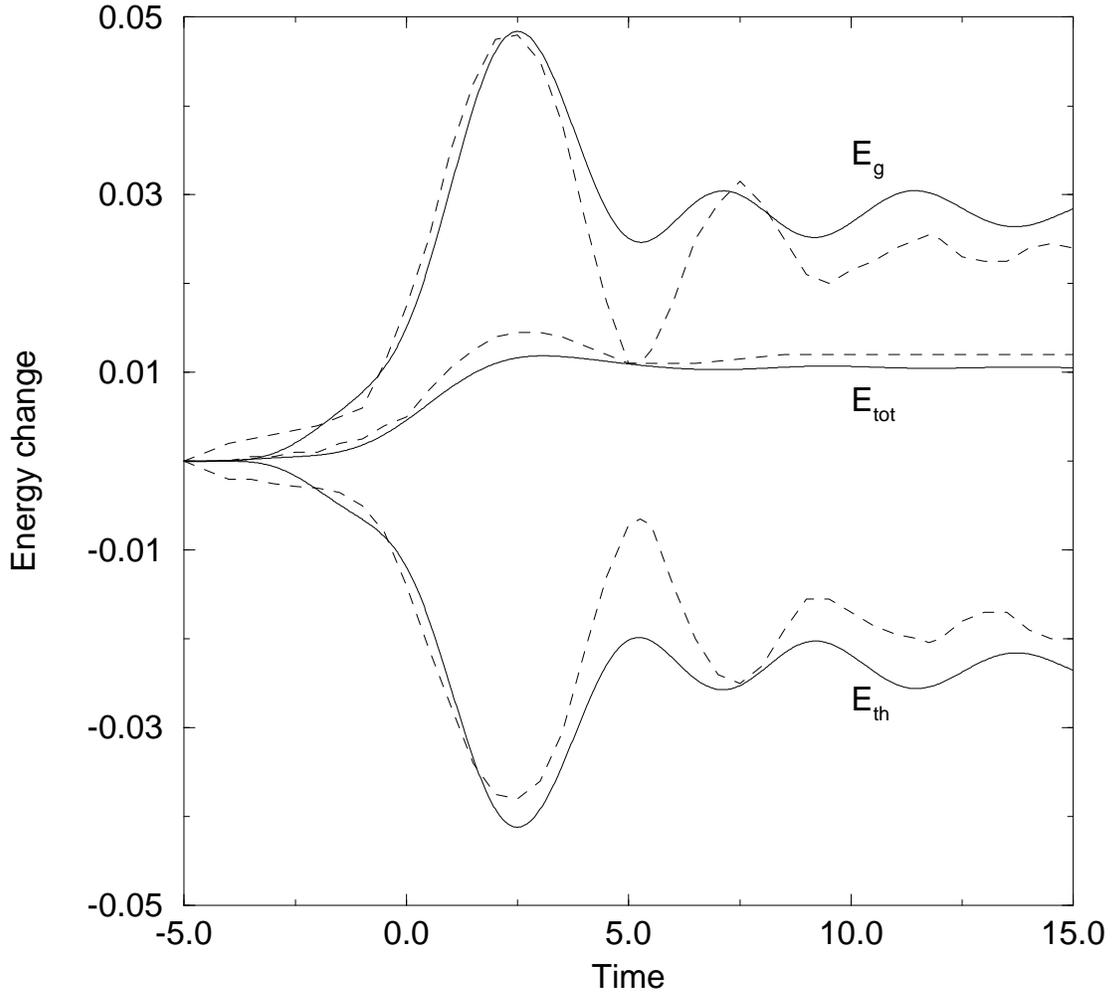}
\caption
{The evolution of the total $E_{tot}$, 
thermal $E_{th}$ and gravitational $E_{g}$
energies
with time. 
All sorts of energy are measured with
respect to their equilibrium values. The solid curves
correspond to our model, 
and the dashed curves correspond to the simulations of Kh a. The case $n=1.5$,
$\eta=3$ is shown.
\label{fig6}}
\end{figure}

\begin{figure}
\plotone{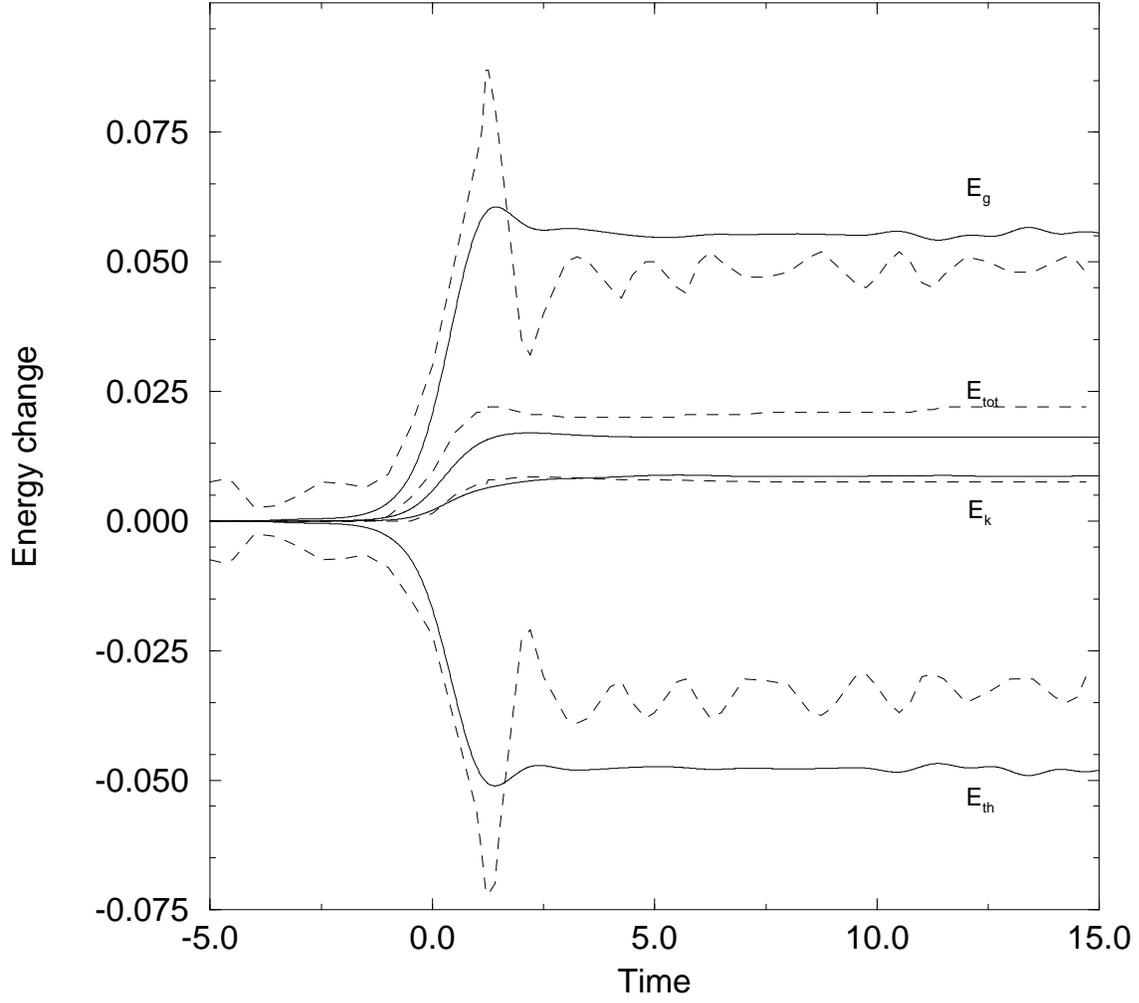}
\caption
{The same as Fig. 6, but $n=3$, $\eta=1.5$. Additionally the evolution 
of the kinetic energy $E_{K}$ is shown.
\label{fig 7}}
\end{figure}
\begin{figure}
\plotone{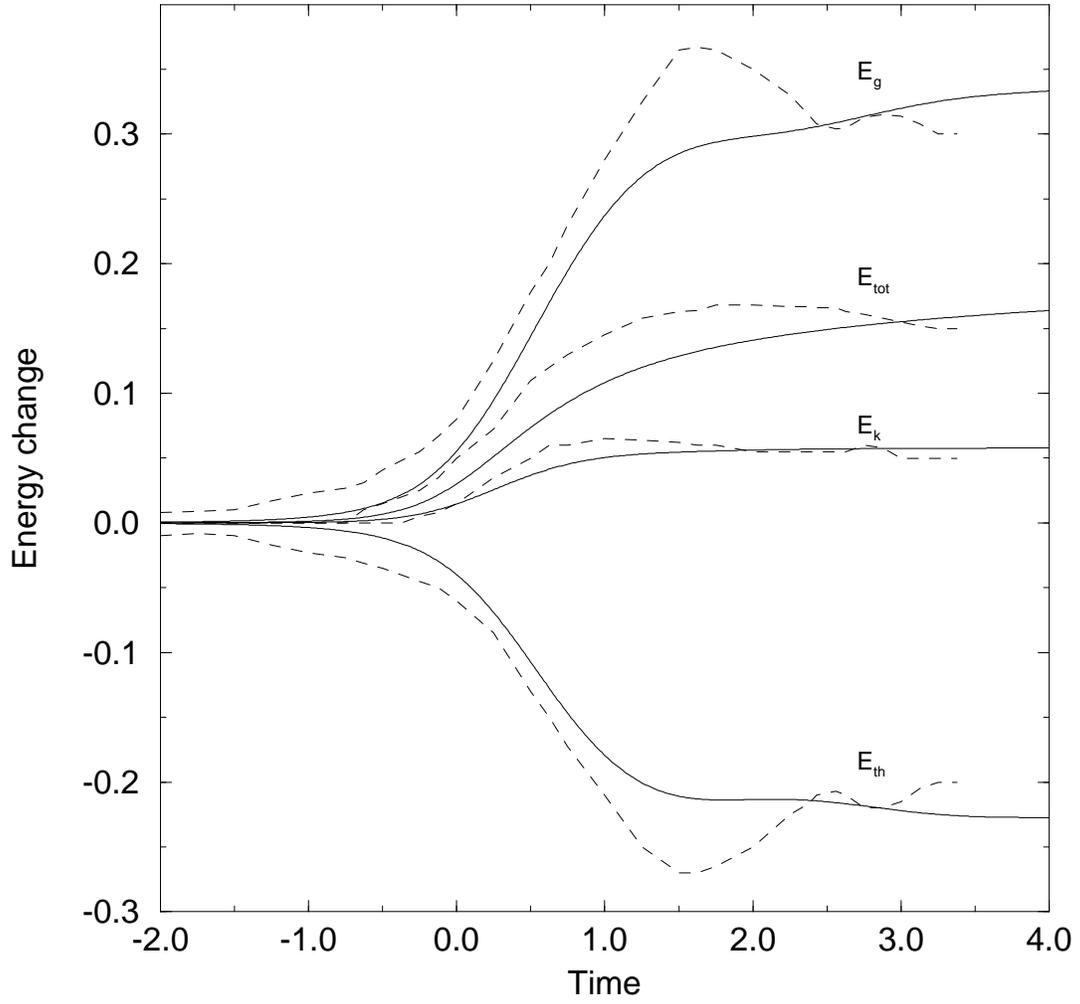}
\caption
{The same as Fig. 7, but $n=3$, $\eta=1$.
\label{fig8}}
\end{figure}
\begin{figure}
\plotone{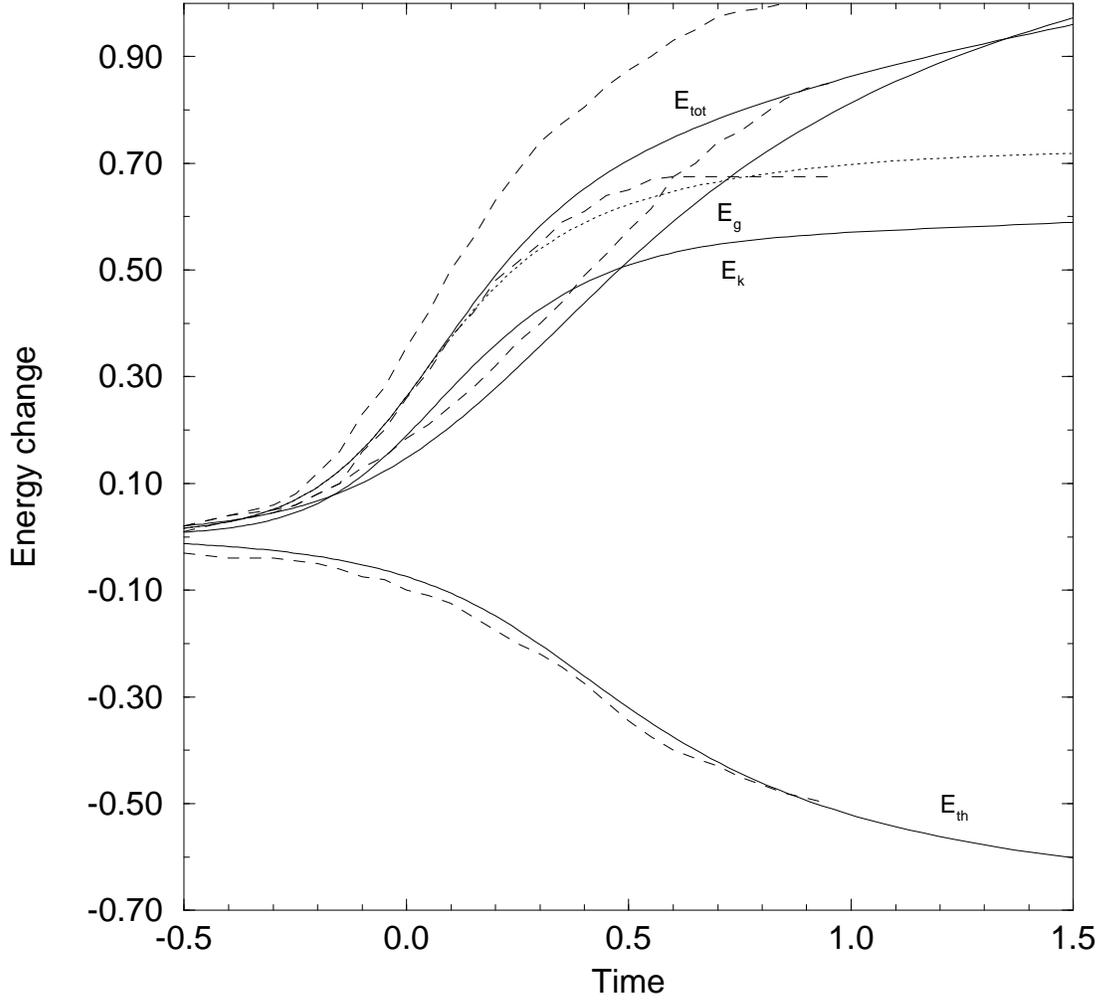}
\caption
{The same as Fig. 7, but $n=3$, $\eta=0.5$. The dashed curves are 
taken from Kh b. Additionally the total energy of the gravitationally bounded part of the
star is shown (dotted curve).
\label{fig9}}
\end{figure}
\begin{figure}
\plotone{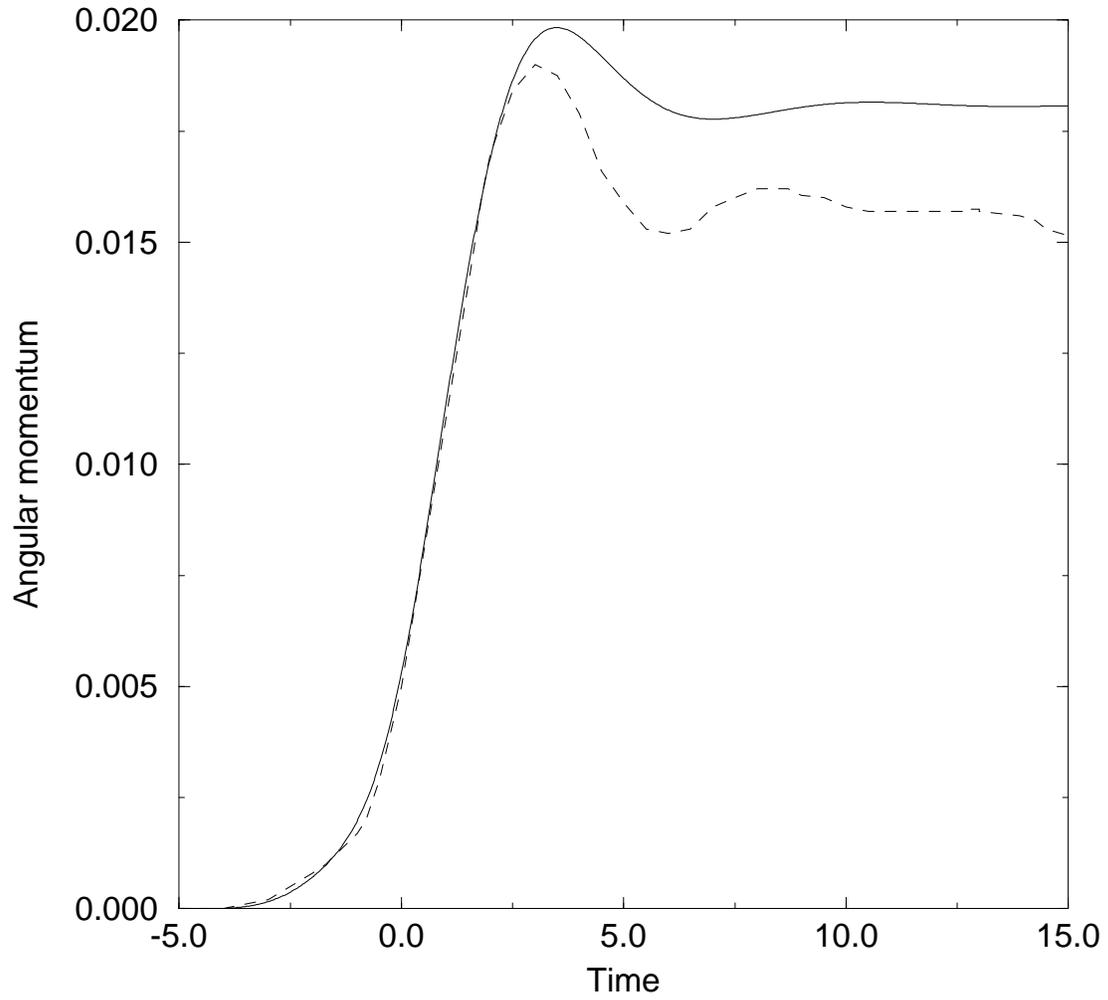}
\caption
{The evolution of $z$ component of angular momentum with time.
The dashed line is taken from Kh a. The case $n=1.5$, $\eta=3$ is shown. 
\label{fig10} }
\end{figure}
\begin{figure}
\plotone{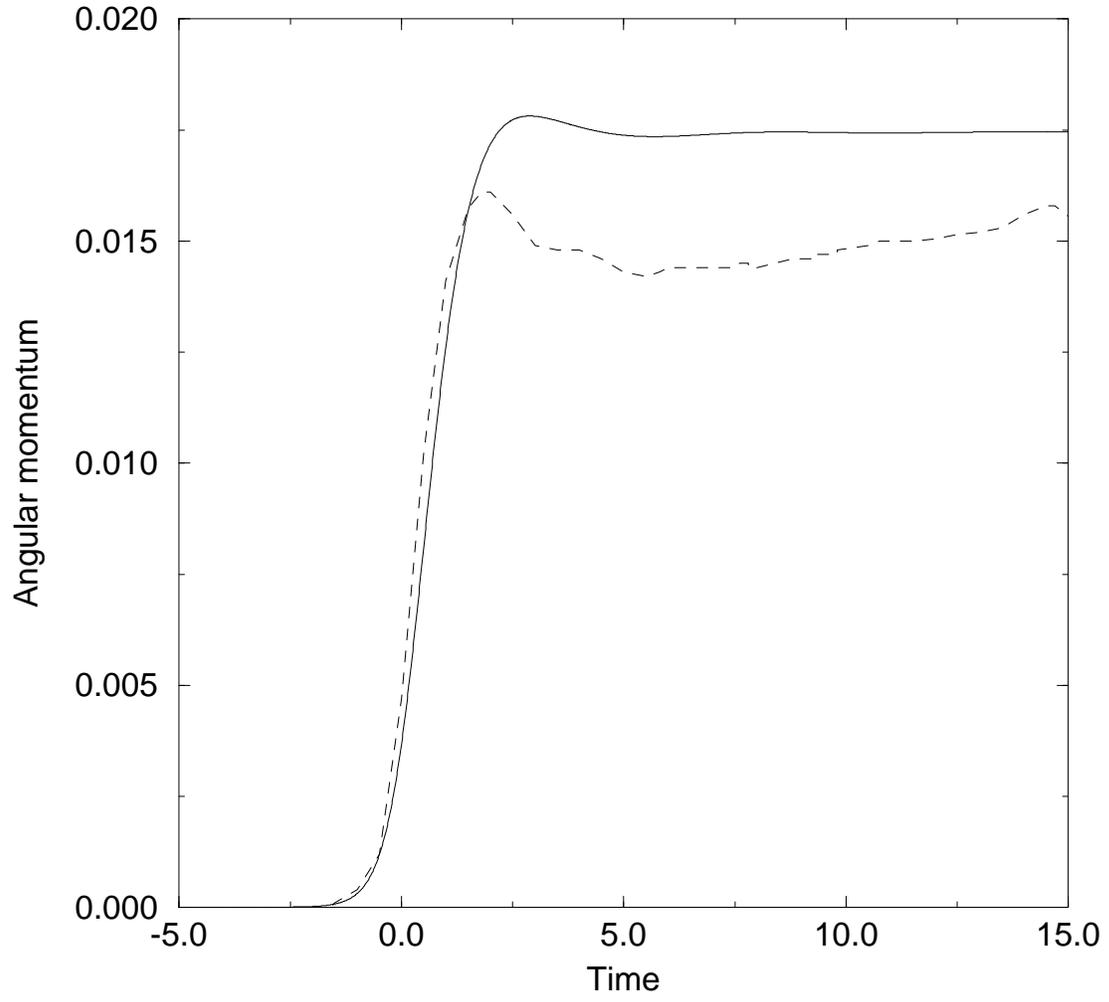}
\caption
{The same as Fig. 10, but $n=3$, $\eta=1.5$.
\label{fig11} }
\end{figure}
\begin{figure}
\plotone{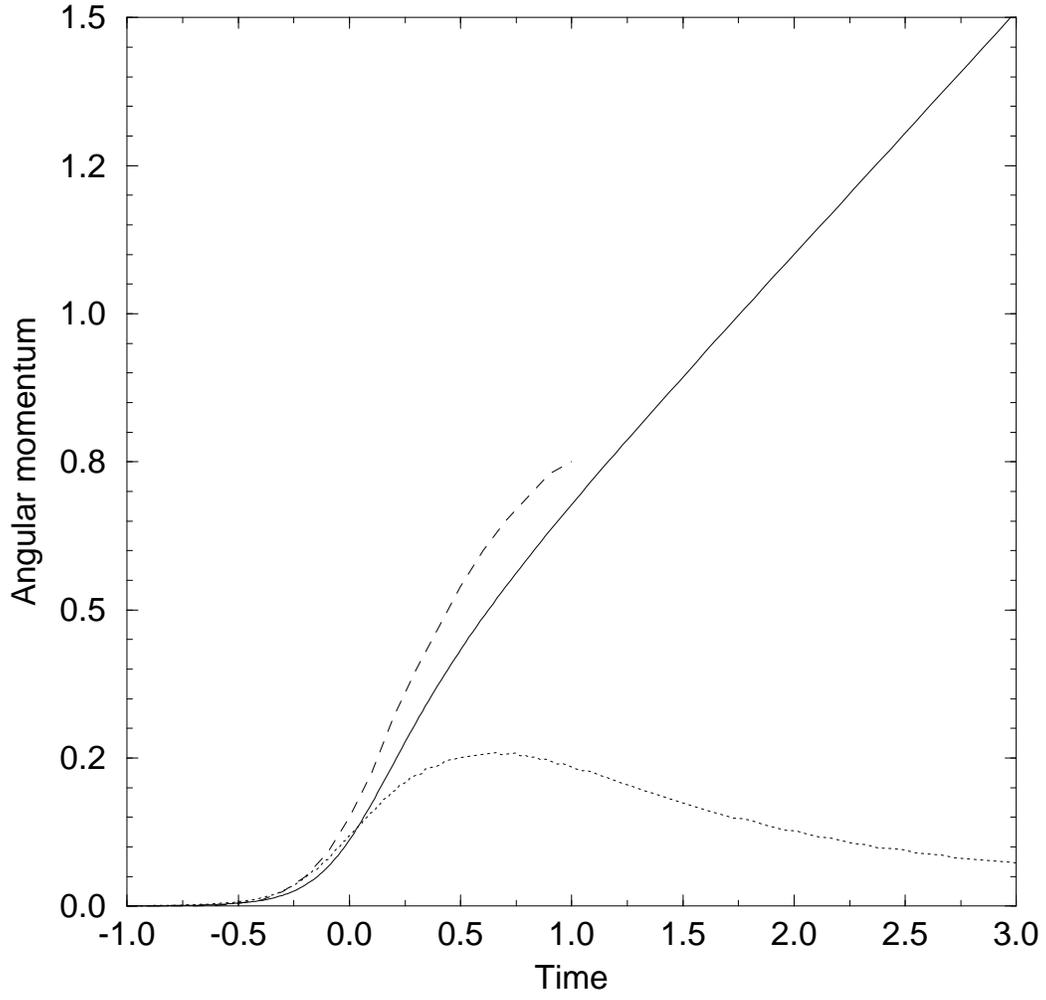}
\caption
{The same as Fig. 10, but $n=3$, $\eta=0.5$.
The dashed line is taken from Kh b. Additionally the angular momentum of  
the gravitationally bounded part of the star is shown (dotted curve).
\label{fig12} }
\end{figure}
\begin{figure}
\plotone{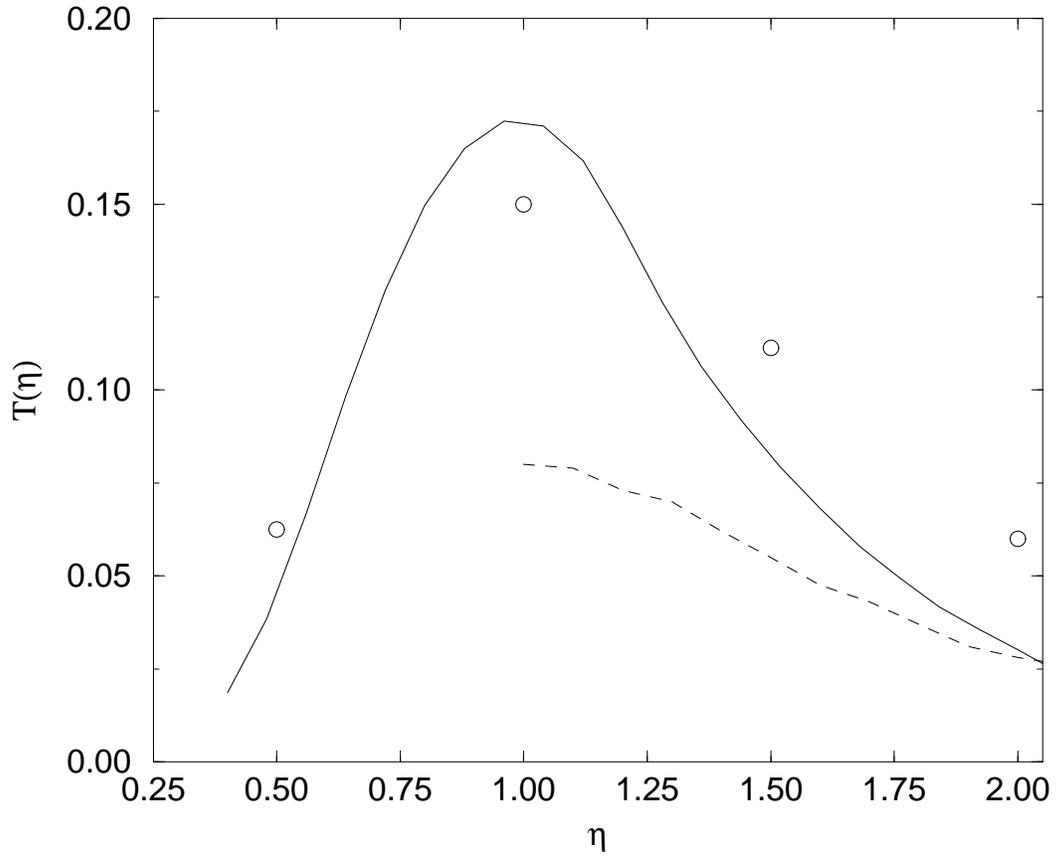}
\caption
{The function $T(\eta)$ is shown for the $n=3$ polytrope.
The dashed curve is taken
from the paper by Lee $\&$ Ostriker 1986 and corresponds to the analytical linear theory.
The open circles show the same quantity
calculated by Kh a, b.  
\label{fig13} }
\end{figure}
\begin{figure}
\plotone{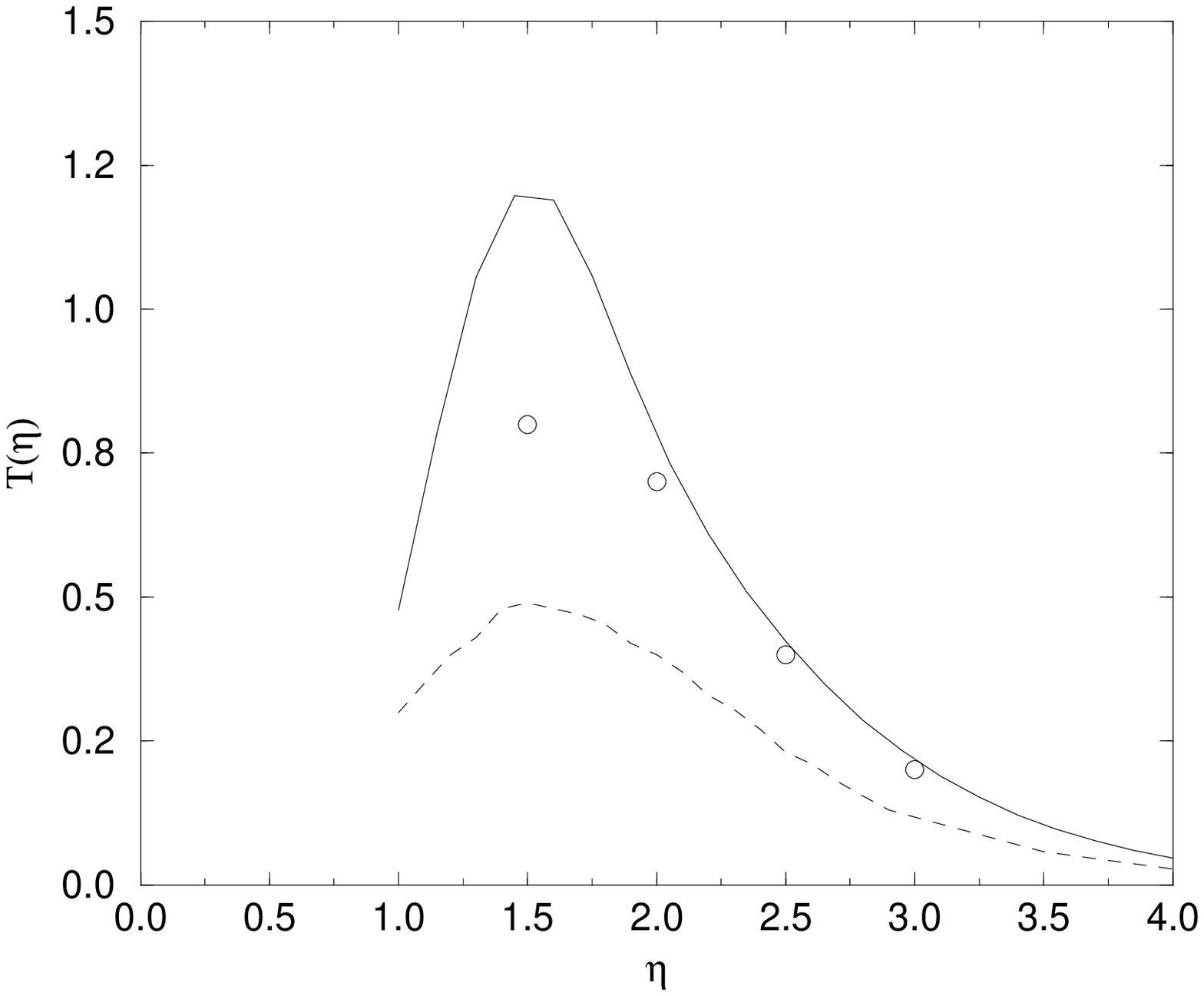}
\caption
{The same as Fig. 13, but for $n=2$ polytrope.
\label{fig14} }
\end{figure}
\begin{figure}
\plotone{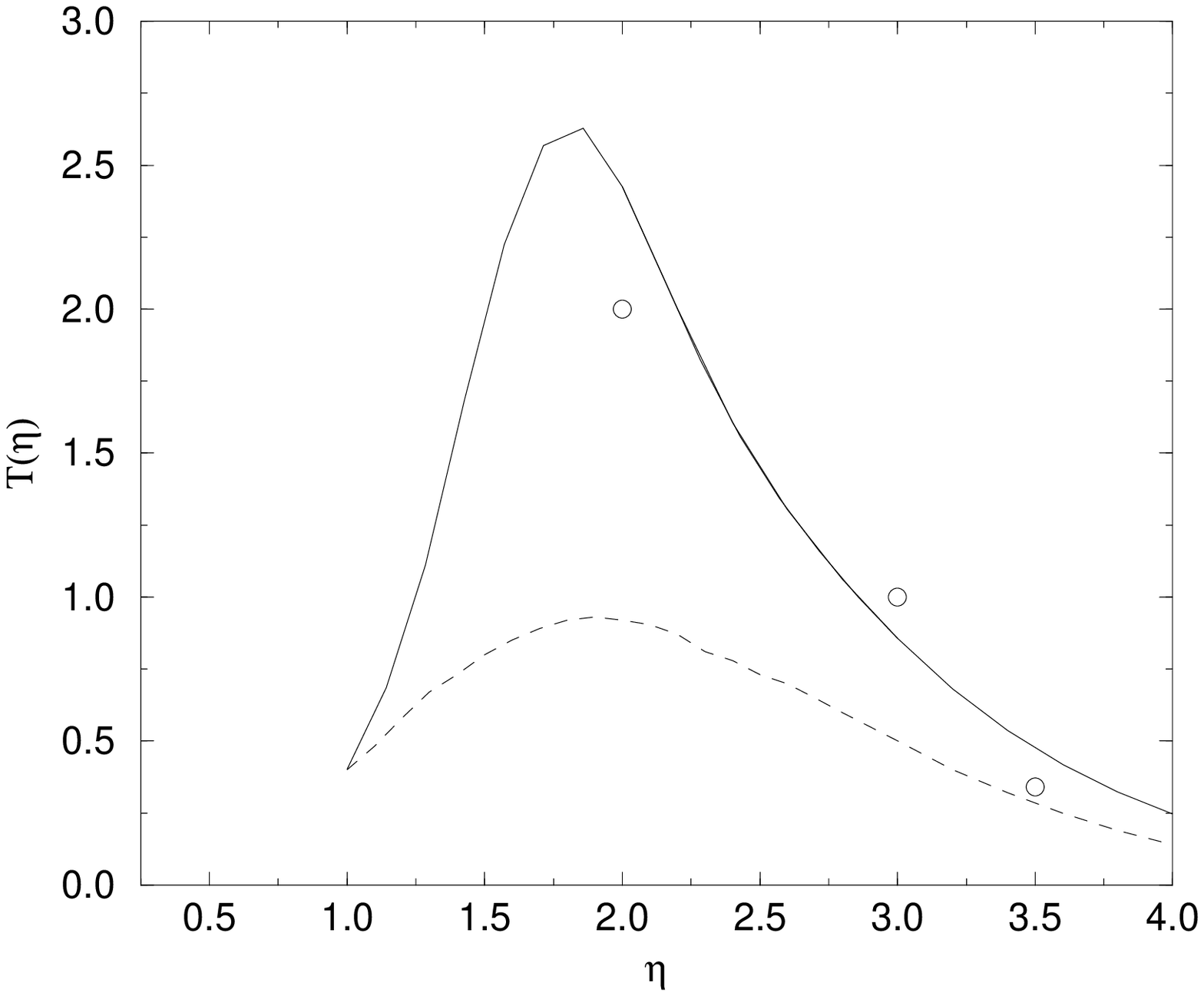}
\caption
{The same as Fig. 13, but for $n=1.5$ polytrope.
\label{fig15} }
\end{figure}
\begin{figure}
\plotone{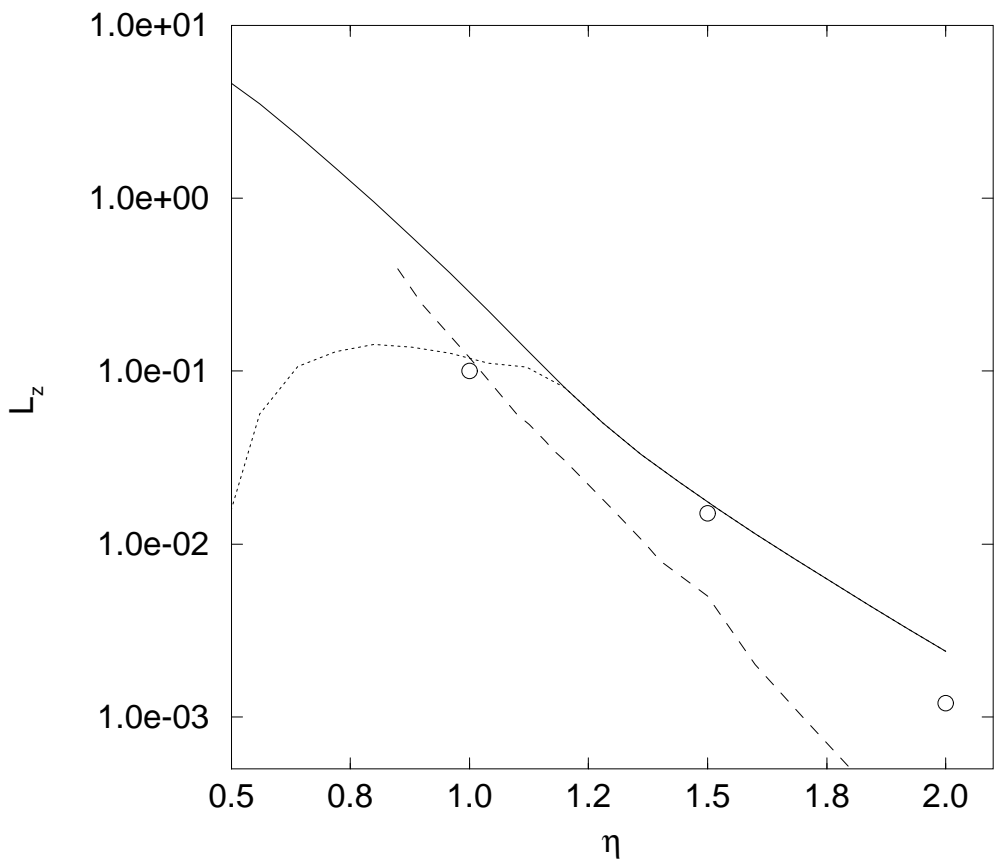}
\caption
{The dependence of the angular momentum gained by the star at the
end of calculations ($\tau_{end}=10$) on the parameter $\eta$. The solid curve represents the
total angular momentum, the dotted line represents the angular momentum of the gravitationally
bounded part of the star. The dashed curve represents the same quantity calculated in the
affine model by Diener et al 1995, and the open circles correspond to the simulations of Kh a, b.
The case of $n=3$ polytrope is shown.
\label{fig16} }
\end{figure}
\begin{figure}
\plotone{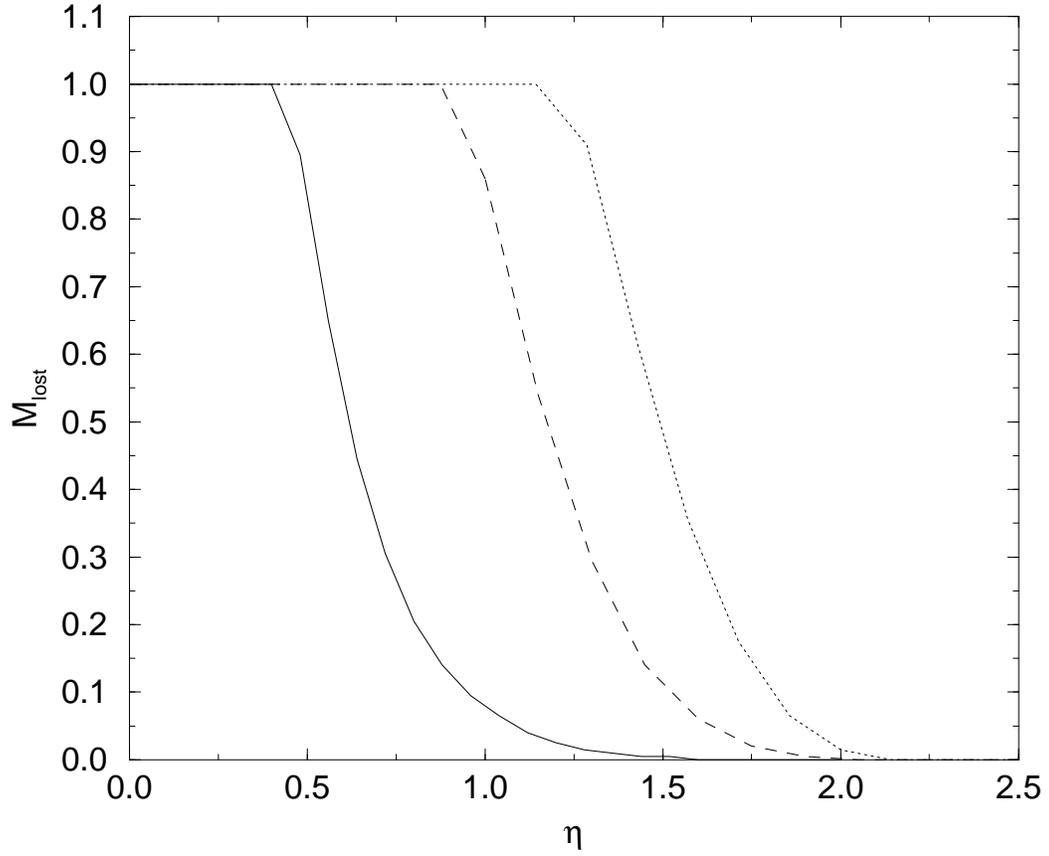}
\caption
{The amount of mass lost by the star in course of tidal encounters.
The solid curve corresponds to the $n=3$ polytrope, the dashed curve corresponds to the
$n=2$ polytrope and the dotted curve corresponds to the
$n=1.5$ polytrope.
\label{fig17} }
\end{figure}





\clearpage



\begin{thebibliography}{}
\bibitem[Auri\`ere(1982)]{aur82}

Ayal, S., Livio, M $\&$ Piran, T. 2000, astro-ph/0002499

\bibitem[Canizares et al.(1978)]{can78}
 
Beloborodov, A. M., Illarionov, A. F., Ivanov, P. B. $\&$ Polnarev, A. G.
1992, \mnras, 259, 209   
 
\bibitem[Djorgovski and King(1984)]{djo84}
 
Bicknell, G. V. $\&$ Gingold, R. A. 1983, \apj, 273, 749

\bibitem[Hagiwara and Zeppenfeld(1986)]{hag86}
 
Cannizzo, J. K., Lee, H. M. $\&$ Goodman, J. 1990, \apj, 351, 38
\bibitem[Harris and van den Bergh(1984)]{har84}
 
Carter, B. $\&$ Luminet, J. P. 1982, \nat, 296, 211

\bibitem[H\`enon(1961)]{hen61}
 
Carter, B. $\&$ Luminet, J.-P. 1983, \aap, 121, 97

\bibitem[King(1966)]{kin66}
 
Carter, B. $\&$ Luminet, J.-P. 1985, \mnras, 212, 23

\bibitem[King(1975)]{kin75}
  
Chandrasekhar, S. 1969, "Ellipsoidal Figures of Equilibrium", New Haven
and London, Yale University Press

\bibitem[King(1975)]{kin75}

Cox, J. P. 1980 "Theory of stellar pulsation", Princeton, New Jersey,
Princeton University Press
   
\bibitem[King et al.(1968)]{kin68}
  
Diener, P., Kosovichev, A. G., Kotok, E. V., Novikov, I. D. $\&$ Pethick,
C. J. 1995, \mnras, 275, 498
    
\bibitem[Kron et al.(1984)]{kro84}

Diener, P., Frolov, V. P., Khokhlov, A. M., Novikov, I. D., $\&$  Pethick,
C. J. 1997 \apj, 479, 164 
   
\bibitem[Lynden-Bell and Wood(1968)]{lyn68} 
   
Dokuchaev, V. I. 1991, \mnras, 251, 564

\bibitem[Newell and O'Neil(1978)]{new78}
   
Evans, C. R. $\&$ Kochanek, C. S 1989, \apjl, 346L, 13

\bibitem[Ortolani et al.(1985)]{ort85}

Fishbone, L. 1973, \apj, 185, 43 
  
\bibitem[Peterson(1976)]{pet76}

Frank, J. $\&$ Rees, M. J. 1976, \mnras, 176, 633

\bibitem[Spitzer(1985)]{spi85}

Frank, J. 1979, \mnras, 187, 883

\bibitem[Spitzer(1985)]{spi85}

Frolov, V. P., Khokhlov, A. M., Novikov, I. D. $\&$ Pethick,
C. J. 1994, \apj,  432, 680

\bibitem[Spitzer(1985)]{spi85}

Fulbright, M. S., Benz, W. $\&$ Davies, M. B 1995, 
\apj, 440, 254

\bibitem[Spitzer(1985)]{spi85}

Gurzadian, V. G.$\&$ Ozernoi, L. M., 1980, \aap, 86, 315

\bibitem[Spitzer(1985)]{spi85}

Hills, J. G. 1975, \nat, 254, 295
\bibitem[Spitzer(1985)]{spi85}

Hills, J. G. 1978, 
\mnras, 182, 517

\bibitem[Spitzer(1985)]{spi85}

Illarionov, A. F. $\&$ Romanova, M. A., 1986a, Soviet Astr., 32, 148
\bibitem[Spitzer(1985)]{spi85}

Illarionov, A. F. $\&$ Romanova, M. A., 1986b, Soviet Astr., 32, 356
\bibitem[Spitzer(1985)]{spi85}

Khokhlov, A. M. 1991, 
\aap,  245, 114

\bibitem[Spitzer(1985)]{spi85}

Khokhlov, A., Novikov, I., D. $\&$ Pethick,
C. J. 1993a, \apj, 418, 163

\bibitem[Spitzer(1985)]{spi85}

Khokhlov, A. M., Novikov, I. D. $\&$ Pethick, C. J. 1993b, \apj, 418, 181
 
\bibitem[Spitzer(1985)]{spi85}

Kim, S. S., Park, M.-G., $\&$ Lee, H. M. 1999, 
\apj, 519, 647

\bibitem[Spitzer(1985)]{spi85}

Kochanek, C., S. 1994, \apj, 423, 344


\bibitem[Spitzer(1985)]{spi85}

Kosovichev, A. G. $\&$ Novikov, I. D. 1992, 
\mnras, 258, 715

\bibitem[Spitzer(1985)]{spi85}

Lacy,J. H., Townes, C. H. $\&$ Hollenbach, D. J. 1982, 
\apj, 262, 120

\bibitem[Spitzer(1985)]{spi85}

Laguna, P., Miller, W. A., Zurek, W. H. $\&$ Davies, M. B. 1993,
\apjl, 410L, 83
\bibitem[Spitzer(1985)]{spi85}
Laguna, P. 1994, 
\apj, 404, 678

\bibitem[Spitzer(1985)]{spi85}

Lai, D., Rasio, F., A. $\&$ Shapiro, S., L. 1994, \apj, 437, 742
\bibitem[Spitzer(1985)]{spi85}

Lai, D. $\&$ Shapiro, S., L. 1995, \mnras, 275, 498

\bibitem[Spitzer(1985)]{spi85}

Lattimer, J. M. $\&$ Schramm, D. N. 1976 \apj, 210, 549

\bibitem[Spitzer(1985)]{spi85}

Lee, H. M., Ostriker, J. P. 1986, \apj, 310, 176

\bibitem[Spitzer(1985)]{spi85}

Luminet, J.-P.$\&$ Pichon, B. 1989, \aap, 209, 103

\bibitem[Spitzer(1985)]{spi85}

Luminet, J.-P. $\&$ Carter, B. 1986, \apjs, 61, 219

\bibitem[Spitzer(1985)]{spi85}

Luminet, J.-P.$\&$ Mark, J.-A. 1985, \mnras, 212, 57

\bibitem[Spitzer(1985)]{spi85}

Magorrian, J. $\&$ Tremaine, S. 1999, \mnras, 309, 477

\bibitem[Spitzer(1985)]{spi85}

Mark, J.-A., Lioure, A. $\&$ Bonazzola, S. 1996, \aap, 306, 666

\bibitem[Spitzer(1985)]{spi85}

Mashhoon, B. 1975, \apj, 197, 705

\bibitem[Spitzer(1985)]{spi85}

Nduka, A. 1971, \apj 170, 131

\bibitem[Spitzer(1985)]{spi85}

Nolthenius, R. A. $\&$ Katz, J. I. 1983, \apj, 269, 297

\bibitem[Spitzer(1985)]{spi85}

Novikov, I. D., Pethick, C. J. $\&$ Polnarev, A. G. 1992, \mnras, 255, 276

\bibitem[Spitzer(1985)]{spi85}

Press, W. H., Teukolsky, S. A. 1977, \apj, 213, 183

\bibitem[Spitzer(1985)]{spi85}

Rees, M. J., 1988, \nat, 333, 523

\bibitem[Spitzer(1985)]{spi85}

Roos, N. 1992, \apj, 385, 108

\bibitem[Spitzer(1985)]{spi85}

Richtmyer, R. D $\&$ Morton, K. W. 1967, ``Difference Methods for Initial-Value
Problems''. Second Edition. New York/ London,
Interscience Publishers

\bibitem[Spitzer(1985)]{spi85}

Syer, D $\&$ Ulmer, A. 1999, \mnras, 306, 35

\bibitem[Spitzer(1985)]{spi85}

Ulmer, A. 1999, \apj, 514, 180

\bibitem[Spitzer(1985)]{spi85}

Ulmer, A., Paczynski, B. $\&$ Goodman, J. 1998, \aap, 333, 379

\bibitem[Spitzer(1985)]{spi85}

Young, P. J., Shields, G. A. $\&$ Wheeler, J. C. 1977, \apj, 212, 367

\bibitem[Spitzer(1985)]{spi85}

Young, P. J. 1977, \apj, 215, 36

\end{thebibliography}
\end{document}